\documentclass[%
 aip,
 jmp,
amsmath,amssymb,
reprint,%
]{revtex4-1}

\usepackage{graphicx}

\usepackage{color} %deactivate me after review
\usepackage{bm}
\usepackage[utf8]{inputenc}
\usepackage[T1]{fontenc}
\usepackage{mathptmx}
\usepackage{soul} %deactivate me after review

\usepackage{cancel}

\usepackage{hyperref}

\newcommand{\ie}{\textit{i.e.}}

\newcommand{\rmd}{\mathrm{d}}
\newcommand{\rmi}{\mathrm{i}}
\newcommand{\rme}{\mathrm{e}}
\newcommand{\tr}{\mathrm{tr}}

\newcommand{\la}{\langle}
\newcommand{\ra}{\rangle}

\newcommand{\One}{\openone}

\newcommand{\h}{\hat}
\newcommand{\rmw}{\mathrm{w}}

\newcommand{\bea}{\begin{eqnarray}}
\newcommand{\eea}{\end{eqnarray}}

\newcommand{\bA}{{\bf A}}

\newcommand{\bPh}{\Phi}

\newcommand{\mc}[1]{\mathcal{#1}}

\begin{document}

%\preprint{AIP/123-QED}

\title[Modulation and protection of entanglement in a two-qubit system via two interacting oscillator control]
{Modulation and protection of entanglement in a two-qubit system via two interacting oscillator control}

\author{P. C. L\' opez V\' azquez}
\email{pablo.lopez@academicos.udg.mx}
\affiliation{Departamento de Ciencias Naturales y Exactas, Universidad de
Guadalajara,
Carretera Guadalajara - Ameca Km. 45.5 C.P. 46600. Ameca, Jalisco, M\'exico.}

\author{A. García}
\affiliation{Departamento de Matemáticas, Universidad de Guadalajara,
Boulevard Marcelino Garc\'ia Barragan y Calzada Ol\'impica, C.P. 44840,
Guadalajara, Jalisco, M\'exico}

\author{J. Romero-Labias}
\affiliation{Departamento de Ciencias Naturales y Exactas, Universidad de
Guadalajara,
Carretera Guadalajara - Ameca Km. 45.5 C.P. 46600. Ameca, Jalisco, M\'exico.}

\date{\today}

\begin{abstract}
We investigate the control and protection of two-qubit entanglement through engineered interactions 
between two auxiliary harmonic oscillators. Each qubit is coupled to a local oscillator via a dephasing 
interaction, while the oscillators interact through independently tunable position-position and 
momentum-momentum couplings. Employing the characteristic-function representation, we obtain 
analytical solutions for the composite-system dynamics and examine how the stability properties 
of the coupled oscillators determine the evolution of two-qubit concurrence. 
Our analysis reveals distinct dynamical regimes exhibiting entanglement revivals, 
suppression, or decay, and identifies configurations in which oscillator interactions effectively 
reduce entanglement degradation. We further investigate the influence of thermal dissipation 
using local and global master-equation descriptions, demonstrating that suitable oscillator 
couplings can preserve substantial two-qubit entanglement under the dissipative conditions considered. 
These results establish a connection between the dynamical stability of auxiliary bosonic modes 
and the controllability of entanglement in discrete-variable quantum systems, providing a potential 
strategy for entanglement protection through engineered oscillator interactions.
\end{abstract}

\maketitle

\section{\label{intro}Introduction}

Entanglement is a fundamental resource in quantum information science, underpinning applications in quantum computation,
quantum communication, and quantum cryptography \cite{nielsen_chuang}. However, its susceptibility to
environmental interactions represents a major obstacle to the reliable implementation of quantum information protocols,
as decoherence can lead to the degradation or loss of quantum correlations \cite{zurek_decoherence,YuEberly2009}.
Consequently, developing strategies to manipulate and protect entanglement remains an important challenge in
quantum technologies. In physical platforms such as superconducting circuits, trapped ions, and hybrid quantum
systems, qubits can interact with auxiliary bosonic degrees of freedom, including electromagnetic and mechanical
oscillators \cite{Blais2021,Aspelmeyer2014}.
Such interactions can induce nontrivial entanglement dynamics, including decay and revivals associated with
the exchange of quantum correlations between different subsystems \cite{YuEberly2009,Breuer2016}.

Several approaches have been developed to mitigate environmentally induced decoherence and protect quantum
information, including dynamical decoupling and quantum error correction~\cite{ViolaLloyd1998,Shor1995}. 
An alternative strategy consists of engineering the interactions between a quantum
system and auxiliary degrees of freedom, exploiting their collective dynamics to influence the evolution of quantum
correlations. In this context, structured environments and coupled bosonic modes provide a particularly
interesting setting, as their internal interactions can modify the effective dynamics experienced by the system
of interest \cite{Breuer2016,Blais2021}.

In a recent work \cite{Lopez26}, we investigated a composite system consisting of two qubits independently
coupled to two harmonic oscillators through dephasing interactions. It was shown that the entanglement dynamics
exhibits a redistribution of quantum correlations among different partitions of the composite system, suggesting
a flux-like behavior associated with the exchange of entanglement between the qubit and oscillator degrees of
freedom. This observation motivates the possibility of controlling such redistribution by modifying the internal
dynamics of the oscillator subsystem. In particular, introducing a tunable interaction between the oscillators
provides an additional degree of control over the evolution of the two-qubit entanglement.

Motivated by these considerations, in the present work we extend the model of Ref.~\cite{Lopez26} by introducing
direct interactions between the two oscillators through independently tunable position--position and
momentum--momentum couplings. These interactions give rise to distinct dynamical regimes, characterized by the
stability properties of the coupled-oscillator system.
We investigate how these regimes influence the entanglement evolution of the two-qubit subsystem,
particularly the occurrence of entanglement revivals, suppression, and decay. Our analysis aims to identify
configurations in which the collective oscillator dynamics effectively reduces the degradation of
two-qubit entanglement.

An important feature of the model is that its quadratic oscillator Hamiltonian and dephasing
interactions allow the dynamics to be treated analytically within the characteristic-function 
representation~\cite{Lopez18,Lopez20,Lopez21}.
This framework enables us to establish a connection between the spectral properties of the oscillator
dynamics and the evolution of quantum correlations in the two-qubit subsystem. Furthermore, interactions
between bosonic modes and dispersively coupled qubits can be engineered in platforms such as circuit quantum
electrodynamics and hybrid quantum devices \cite{Blais2021,Gao2018,Zhang2019,Aspelmeyer2014}, 
suggesting possible physical settings in which related control mechanisms could be investigated.

Beyond the unitary dynamics, we address the robustness of the proposed entanglement-control mechanism in the
presence of environmental dissipation. To this end, we consider two descriptions of the oscillator--environment
interaction. The first is a local dissipative model, in which each oscillator is coupled to an independent
thermal reservoir. The second is a global master-equation description, in which dissipation is formulated in
terms of the collective normal modes of the interacting oscillator system. The latter incorporates the effects
of the internal oscillator coupling into the dissipative dynamics and is particularly relevant when such
interactions cannot be neglected in constructing the master equation. By comparing these two descriptions,
we investigate the extent to which suitable oscillator couplings can enhance the preservation of two-qubit
entanglement under dissipative conditions.

Our results demonstrate that the interaction between auxiliary bosonic modes provides a means of modulating
the entanglement dynamics of a discrete-variable quantum system. In particular, we identify dynamical
regimes in which appropriately chosen oscillator couplings suppress entanglement degradation, both in
the absence and in the presence of dissipation. This establishes a connection between the dynamical properties 
of the auxiliary oscillator subsystem and the possibility of protecting entanglement through engineered 
interactions.

The paper is organized as follows. In Sec.~\ref{model}, we introduce the composite
two-qubit--two-oscillator model and the local and global dissipative descriptions. In
Sec.~\ref{sec:entanglement-dynamics} and \ref{sec:markovian-transition}, 
we examine the influence of the oscillator interactions on the two-qubit
entanglement dynamics and its relationship to the stability properties of the oscillator 
subsystem. In Sec.~\ref{sec:local-dissipation} and \ref{sec:global-dissipation}, 
we analyze entanglement protection in the presence of dissipation.
Finally, in Sec.~\ref{conclusion}, we summarize
our main findings and discuss their implications.

\section{\label{model} The model}
\begin{figure}[!htbp]
  \begin{center}
  \includegraphics[scale=0.5]{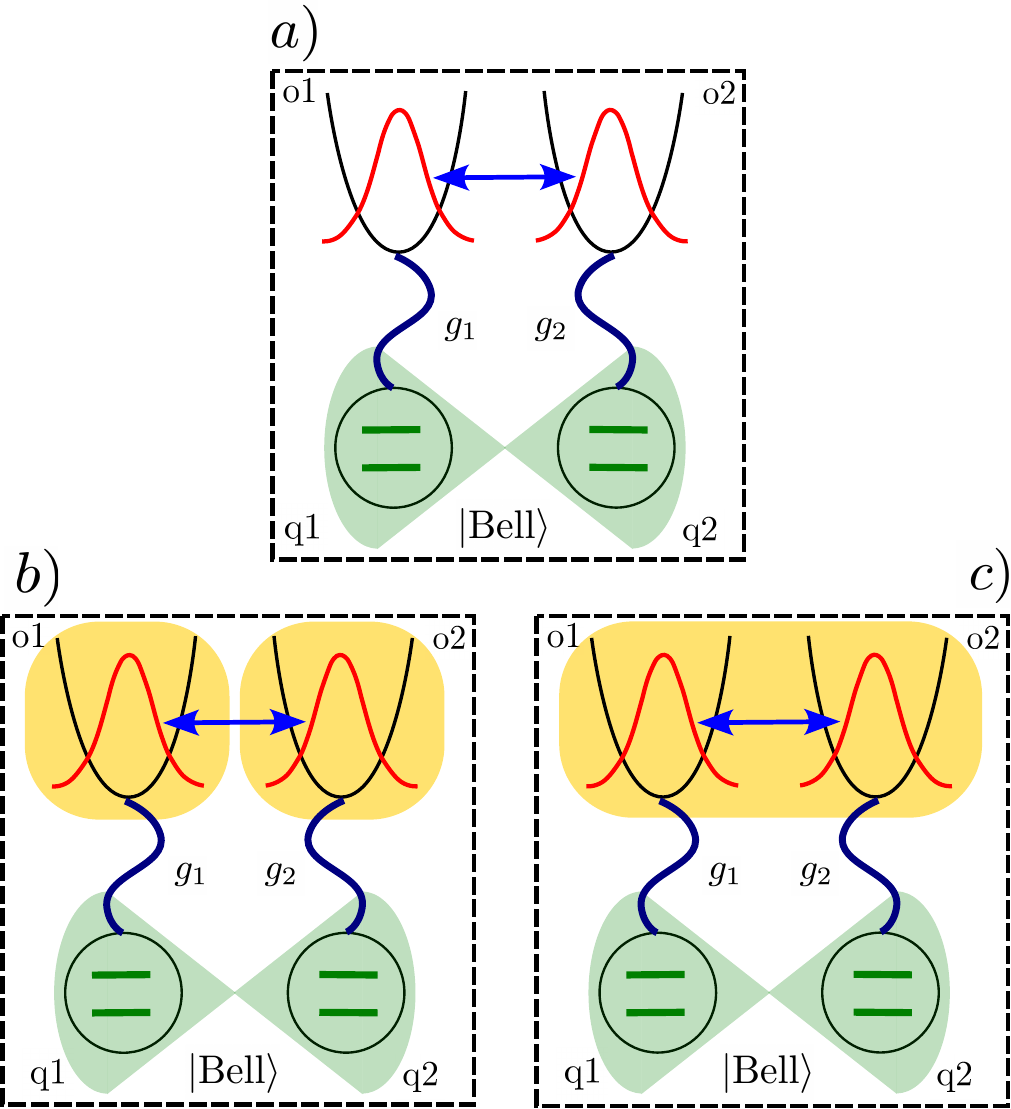}
  \end{center}
  \caption{\label{fig1}
Schematic representation of the composite system consisting of two qubits coupled to two interacting
harmonic oscillators. Each qubit interacts with its corresponding oscillator through a local dephasing 
interaction, while the oscillators are coupled through tunable position--position and momentum--momentum 
interactions.
Panel (a) illustrates a configuration where no dissipation is present and the qubits are initially prepared in a
maximally entangled Bell state.
Panel (b) depicts the configuration investigated in the entanglement-protection analysis, where
the qubits are initially prepared in a maximally entangled Bell state and the oscillator subsystem is 
subject to local dissipation.
Panel (c) also depicts the configuration investigated in the entanglement-protection analysis, where
the qubits are initially prepared in a maximally entangled Bell state but now the oscillator 
subsystem is subject to global dissipation.}
\end{figure}

We consider a composite system consisting of two qubits, each coupled to a harmonic oscillator through a local
dephasing interaction. In addition, the two oscillators interact through independently tunable position--position
and momentum--momentum couplings. We first investigate the dynamics of the composite system in the absence of
dissipation and subsequently incorporate the influence of thermal environments acting on the oscillator degrees
of freedom. Two dissipative descriptions are considered, corresponding to local and global master equations.
A schematic representation of the system is shown in Fig.~\ref{fig1}

The Hamiltonian of the composite system is given by
\begin{equation}
H=H_{\mathrm{qubits}}+H_{\mathrm{oscillators}}
  +H_{\mathrm{coupling}}
  +H_{\mathrm{oscillator\_interaction}},
\end{equation}
where
\begin{equation}
H_{\mathrm{qubits}}
= \frac{\Delta_1}{2}\sigma_z^{(1)} + \frac{\Delta_2}{2}\sigma_z^{(2)},
\end{equation}
\begin{equation}\label{hoscillators}
H_{\mathrm{oscillators}}
=\omega_a\h{n}_a+\omega_b\h{n}_b,
\end{equation}
\begin{equation}
H_{\mathrm{coupling}}
= g_1\sigma_z^{(1)}\h{x}_a + g_2\sigma_z^{(2)}\h{x}_b,
\end{equation}
and
\begin{equation}\label{hinteractionoscillators}
H_{\mathrm{oscillator\_interaction}}
= \alpha\h{x}_a\h{x}_b + \beta\h{p}_a\h{p}_b.
\end{equation}

Here, $\sigma_z^{(i)}$ denotes the Pauli $z$ operator of the $i$-th qubit, while $\h{n}_j$,
$\h{x}_j$, and $\h{p}_j$ are the number, position, and momentum operators, respectively,
of oscillator $j=a,b$. We adopt dimensionless quadratures satisfying
$[\h{x}_j,\h{p}_k]=\rmi\delta_{jk}$ and measure all frequencies and interaction strengths
in units of the natural frequency of the first oscillator, $\omega_{\mathrm{o1}}$, setting
$\hbar=1$. Accordingly, $\Delta_i=\omega_{\mathrm{qi}}/\omega_{\mathrm{o1}}$
and $g_i=\lambda_i/\omega_{\mathrm{o1}}$, where $\omega_{\mathrm{qi}}$ and $\lambda_i$
are the qubit frequencies and qubit--oscillator coupling strengths, respectively. The oscillator
frequencies are thus $\omega_a=1$ and
$\omega_b=\Omega=\omega_{\mathrm{o2}}/\omega_{\mathrm{o1}}$, while
$\alpha=\tilde{\alpha}/\omega_{\mathrm{o1}}$ and $\beta=\tilde{\beta}/\omega_{\mathrm{o1}}$
represent the dimensionless oscillator--oscillator coupling strengths.

The parameters $\alpha$ and $\beta$ determine the nature of the interaction between the
oscillators and, consequently, their collective dynamics. In particular, for $\alpha=\beta$,
the interaction reduces to an excitation-exchange Hamiltonian,
\begin{equation}
H_{\mathrm{oscillator\_interaction}}
=\alpha\left( \h{a}_a^\dagger\h{a}_b + \h{a}_a\h{a}_b^\dagger \right),
\end{equation}
whereas $\alpha\neq\beta$ introduces additional two-mode squeezing terms. These interactions
provide a means of modifying the dynamical properties of the oscillator subsystem.

An important aspect of the model is that the oscillator interactions can give rise to distinct
dynamical regimes, depending on the values of
$\alpha$, $\beta$, and $\Omega$; these will be our tunable parameters. These regimes
are determined by the eigenvalues of the corresponding stability matrix, whose explicit
form and spectral properties are presented in Appendix~\ref{app1}. In particular,
purely imaginary eigenvalues characterize oscillatory linear dynamics, whereas eigenvalues
with nonzero real parts indicate dynamically unstable regimes.

It is important to distinguish dynamical stability from the boundedness of the quadratic
oscillator Hamiltonian. For $\Omega>0$, the Hamiltonian is positive definite in its
quadratic degrees of freedom when
\begin{equation}\label{positive_hamiltonian}
|\alpha|<\sqrt{\Omega},
\qquad
|\beta|<\sqrt{\Omega}.
\end{equation}
Within this parameter region, the oscillator Hamiltonian admits a decomposition into
positive-energy normal modes with real, positive frequencies. Outside this region,
oscillatory dynamics may still occur for certain parameter configurations, although
the Hamiltonian need not be bounded from below. This distinction will be relevant when
considering the different dynamical regimes and the construction of the dissipative models.

\subsection*{Dissipative descriptions}

To investigate the influence of environmental dissipation, we describe the evolution of the
density operator $\varrho$ of the full composite system through a Markovian master equation
of the form
\begin{equation}\label{master_eq}
\rmi\dot{\varrho} = [H,\varrho]+\rmi\mathcal{D}[\varrho],
\end{equation}
where $\mathcal{D}[\varrho]$ represents the dissipative contribution associated with
the coupling of the oscillators to external thermal environments. The environments are
assumed to couple directly to the oscillator degrees of freedom, while the qubits are
affected indirectly through their dephasing interactions with the oscillators.

We consider two descriptions of the dissipative dynamics:
\begin{itemize}
\item \textit{Local dissipation.}
In this description, each oscillator is coupled to an independent thermal reservoir.
The dissipative dynamics is modeled through the bare oscillator operators, yielding
\begin{equation}\label{Dlocal}
\mathcal{D}_{\mathrm{local}}[\varrho]
=\sum_{j=a,b}
\left[
\gamma_j(\bar{n}_j+1) \mathcal{D}[\h{a}_j]\varrho
+\gamma_j\bar{n}_j \mathcal{D}[\h{a}_j^\dagger]\varrho
\right],
\end{equation}
where $\gamma_j$ denotes the dissipation rate of oscillator $j$ and $\bar{n}_j$
is the mean thermal occupation of its reservoir. The Lindblad dissipator is defined as
\begin{equation}
\mathcal{D}[\h{O}]\varrho
=\h{O}\varrho\h{O}^\dagger - \frac{1}{2} \left\{\h{O}^\dagger\h{O},\varrho\right\}.
\end{equation}

This local description is commonly employed when the internal oscillator coupling is
sufficiently weak for the use of bare oscillator jump operators to provide an
adequate approximation. Its validity also depends on the assumptions underlying the
Markovian and secular treatments
\cite{Hofer_2017}.

\item \textit{Global dissipation.}
In this description, the dissipative dynamics is formulated in terms of the collective
normal modes of the interacting oscillator subsystem. The corresponding dissipator is
written as
\begin{equation}\label{Dglobal}
\mathcal{D}_{\mathrm{global}}[\varrho]
=\sum_{k=1,2}
\left[
\gamma_k(\bar{n}_k+1)
\mathcal{D}[\h{A}_k]\varrho
+\gamma_k\bar{n}_k
\mathcal{D}[\h{A}_k^\dagger]\varrho
\right],
\end{equation}
where the normal-mode annihilation operators are related to the original oscillator
operators through the Bogoliubov transformation
\begin{equation}\label{Bogoliubov_A}
\h{A}_k = \sum_{j=a,b}
\left( u_{kj}\h{a}_j+ v_{kj}\h{a}_j^\dagger \right).
\end{equation}

The coefficients $u_{kj}$ and $v_{kj}$ are determined by the diagonalization of
the coupled-oscillator Hamiltonian, as described in Appendix~\ref{app3}. 
Here, $\gamma_k$ represents the decay rate of the $k$-th normal mode, while
$\bar{n}_k=\bar{n}(\tilde{\omega}_k)$ denotes its mean thermal occupation,
evaluated at the corresponding normal-mode frequency
$\tilde{\omega}_k =\sqrt{\lambda_k(\alpha)\lambda_k(\beta)}$.
The quantities $\lambda_k(\alpha)$ and $\lambda_k(\beta)$ are defined in
Appendix~\ref{app3}.

Unlike the local description, the global dissipator accounts for the internal
oscillator interaction in the definition of the dissipative modes. Its thermal
normal-mode interpretation is adopted within the parameter region where the
quadratic oscillator Hamiltonian admits a positive-energy diagonalization,
as specified by Eq.~(\ref{positive_hamiltonian}).
The local and global descriptions therefore provide complementary approximations
for investigating the influence of oscillator dissipation on the qubit
entanglement dynamics.

\end{itemize}

\subsection*{Characteristic-function representation}

To obtain the dynamics of the composite system, we introduce the two-qubit
computational basis
$|g_1g_2\rangle=|1\rangle$,
$|g_1e_2\rangle=|2\rangle$,
$|e_1g_2\rangle=|3\rangle$, and
$|e_1e_2\rangle=|4\rangle$,
where $|g_i\rangle$ and $|e_i\rangle$ denote the ground and excited states of
qubit $i$, respectively.

Projecting Eq.~(\ref{master_eq}) onto this basis yields a set of dynamical equations
for the oscillator operators $\varrho_{ij}=\langle i|\varrho|j\rangle$,
which can be formally written as
\begin{equation}\label{densmatrix_elements}
\rmi\dot{\varrho}_{ij}
=\mathcal{L}_{ij}[\varrho]
+\rmi\mathcal{D}_{ij}[\varrho],
\end{equation}
where
$\mathcal{L}_{ij}[\varrho]
=\langle i|[H,\varrho]|j\rangle$
and
$\mathcal{D}_{ij}[\varrho]
=\langle i|\mathcal{D}[\varrho]|j\rangle$.
Owing to the dephasing character of the qubit--oscillator interactions, these
equations do not couple different qubit matrix elements. Their explicit
forms are given in Appendix~\ref{app1}.

The analytical treatment is conveniently carried out using the 
characteristic-function representation of the oscillator degrees
of freedom \cite{Ozo98,Ozo02,Oz04}. Introducing the four-dimensional
Fourier phase-space vector
$\vec{R}=(k_1,s_1,k_2,s_2)^T$,
the corresponding characteristic functions are defined as
\begin{equation}\label{transfw}
w_{ij}(\vec{R},t)
=\int \rmd\vec{q}\,
\varrho_{ij}(\vec{q},\vec{s},t)
\rme^{\rmi\vec{k}\cdot\vec{q}},
\end{equation}
where
$\vec{k}=(k_1,k_2)^T$,
$\vec{s}=(s_1,s_2)^T$,
$\vec{q}=(q_1,q_2)^T$, and

\begin{equation}
\varrho_{ij}(\vec{q},\vec{s},t) = \left\la q_1\!+\!{s_1\over 2}, q_2\!+\!{s_2\over 2}\right| \varrho_{ij}(t)
\left|q_1\!-\!{s_1\over 2}, q_2\!-\!{s_2\over 2}\right\ra .
\end{equation}

We consider initial states that are factorized with respect to the qubit
and oscillator subsystems,
\begin{equation}\label{ic}
\rmw(\vec{R},0)
=\varrho_{\mathrm{q's}}(0)
\,w_{\mathrm{o's}}(\vec{R},0),
\end{equation}
where $\varrho_{\mathrm{q's}}(0)$ is the initial two-qubit density matrix
and $w_{\mathrm{o's}}(\vec{R},0)$ is the characteristic function of the
initial two-oscillator state. In general, the oscillator state may be chosen
according to the physical configuration under investigation.

Under the transformations introduced above, the evolution equations become
first-order partial differential equations in the Fourier phase-space variables.
Their analytical solutions can be expressed in terms of the transition matrix
associated with the oscillator dynamics and, in the dissipative cases, additional
contributions describing diffusion.
The corresponding derivations and explicit solutions are presented in the appendices.

The characteristic-function representation of the full composite state at time $t$
can then be written as
\begin{equation}\label{evtot}
\rmw(\vec{R},t)
=\sum_{i,j=1}^{4}
w_{ij}(\vec{R},t)
|i\rangle\langle j|.
\end{equation}

The reduced dynamics of the two-qubit subsystem follows directly by evaluating
the characteristic function at the origin of Fourier phase space,
\begin{equation}\label{Nonentqrho}
\varrho_{\mathrm{q's}}(t)
=\rmw(\vec{R},t)\big|_{\vec{R}=0}
=\sum_{i,j=1}^{4}
w_{ij}(0,t)|i\rangle\langle j|.
\end{equation}
Similarly, tracing over the qubit degrees of freedom yields the
characteristic function of the reduced two-oscillator subsystem,
\begin{equation}\label{oscdyn}
w_{\mathrm{o's}}(\vec{R},t)
=\sum_{i=1}^{4}w_{ii}(\vec{R},t).
\end{equation}

\subsection*{Two-qubit entanglement dynamics}

Our main interest is to determine how the interaction between the oscillators
influences the entanglement dynamics of the two-qubit subsystem. To this end,
we consider the qubits initially prepared in the maximally entangled Bell state
\begin{equation}
|\Psi^+\rangle
=\frac{1}{\sqrt{2}}
\left(
|e_1g_2\rangle+|g_1e_2\rangle
\right),
\end{equation}
while the oscillator subsystem is initially uncorrelated with the qubits.
Thus, the initial state corresponds to
\begin{equation}
\varrho(0)
=|\Psi^+\rangle\langle\Psi^+|
\otimes\varrho_{\mathrm{o's}}(0).
\end{equation}

Because the interactions preserve the populations of the qubit computational
basis states, the characteristic-function matrix retains the structure
\begin{equation}\label{solentq}
\rmw(\vec{R},t)
=\frac{1}{2}
\begin{pmatrix}
0&0&0&0\\
0&\chi_+(\vec{R},t)&f(\vec{R},t)&0\\
0&f^*(-\vec{R},t)&\chi_-(\vec{R},t)&0\\
0&0&0&0
\end{pmatrix},
\end{equation}
where $\chi_\pm(\vec{R},t)$ describe the evolution of the diagonal
oscillator components and $f(\vec{R},t)$ describes the coherence
between the two qubit states.
Their explicit analytical expressions are given in the appendices for the
corresponding dynamical configurations.

Taking the partial trace over the oscillator subsystem, and using
$\chi_\pm(0,t)=1$, we obtain
\begin{equation}\label{rhobq}
\varrho_{\mathrm{q's}}(t)
=\frac{1}{2}
\begin{pmatrix}
0&0&0&0\\
0&1&f(t)&0\\
0&f^*(t)&1&0\\
0&0&0&0
\end{pmatrix},
\end{equation}
where $f(t)=f(\vec{R}=0,t)$.

The reduced two-qubit density matrix has an $X$-state structure, for which
the concurrence \cite{Wo01,Qu12} takes the particularly simple form
\begin{equation}\label{Eq:q1q2-concurrence}
\mc{C}(t)=|f(t)|.
\end{equation}

Consequently, the entanglement dynamics of the two-qubit subsystem is entirely
determined by the evolution of a single coherence function. In the following
section, we use this result to investigate how the oscillator interaction parameters
$\alpha$ and $\beta$ influence the entanglement dynamics, particularly
through the different stable and unstable regimes of the coupled
oscillator subsystem. We subsequently examine the extent to which these
interactions can protect entanglement in the presence of local and global
dissipation.

%%%%%%%%%%%%%%%%%%%%%%%%%%%%%%%%%%%%%%%%%%%%%%%%%%%%%%%%%%%%%%%%%%%%%%%%%%%%%%%%%%%%%%%%%%%%%%%%%%%%%%%%%%%%%
\section{\label{sec:results} Results}

\subsection{\label{sec:entanglement-dynamics}
Entanglement dynamics in the absence of dissipation}

We begin by examining how the interaction between the two oscillators influences the 
entanglement dynamics of the two-qubit subsystem in the absence
of environmental dissipation. In particular, we investigate the role of the tunable coupling
parameters $\alpha$ and $\beta$, which determine the collective dynamics of the oscillator 
subsystem and, consequently, modify the evolution of the two-qubit concurrence.

As discussed in Appendix~\ref{app1}, the nondissipative dynamics of the coupled oscillators
is characterized by the eigenvalues of the stability matrix~(\ref{stabilityMat}), which
depend on $\alpha$, $\beta$, and the oscillator frequency $\Omega$. Depending on these 
parameters, the eigenvalues may be purely imaginary, giving rise to oscillatory dynamics, 
or possess nonzero real parts, associated with dynamical instabilities.
These distinct regimes provide a natural framework for investigating the influence of the 
oscillator interactions on the two-qubit entanglement.

The parameter space can be classified into the regions illustrated in Fig.~\ref{stab1}:

\begin{itemize}

\item \fbox{\textbf{P0}} corresponds to the
central region $|\alpha|<\sqrt{\Omega}$ and $|\beta|<\sqrt{\Omega}$, where the
oscillator dynamics is stable and the quadratic Hamiltonian is bounded from below.
The particular configuration $\alpha=\beta=0$, corresponding to
uncoupled oscillators, will serve as our reference case.

\item \fbox{\textbf{P1a}} and \fbox{\textbf{P1b}} correspond to dynamically unstable 
regions in which the stability matrix possesses eigenvalues with nonzero real parts, 
leading to exponentially growing oscillator trajectories.

\item \fbox{\textbf{P2a}} and \fbox{\textbf{P2b}} are characterized by purely imaginary 
eigenvalues and therefore exhibit oscillatory dynamics. Unlike the central region P0, 
however, the quadratic oscillator Hamiltonian is not bounded from below in these
regions, despite the absence of exponentially growing trajectories.

\item \fbox{\textbf{P3a}},
\fbox{\textbf{P3b}}, \fbox{\textbf{P3c}}, and \fbox{\textbf{P3d}} correspond to
mixed dynamical regimes, in which the stability matrix possesses both real
and purely imaginary eigenvalues. Consequently, oscillatory and unstable
contributions coexist in the oscillator dynamics.

\end{itemize}

\begin{figure*}[ht]
\centering
  \includegraphics[width=\textwidth]{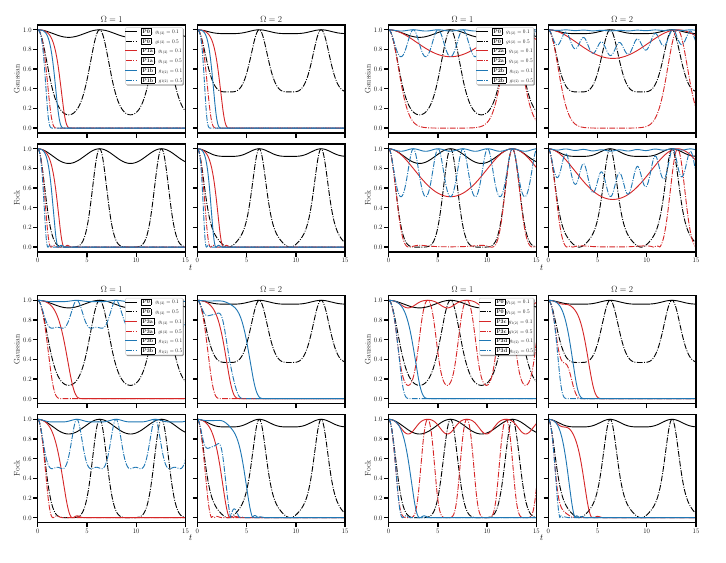}
  \caption{\label{fig2}
Time evolution of the two-qubit concurrence $\mc{C}(t)$ for representative 
oscillator interaction parameters $(\alpha,\beta)$ belonging to the dynamical regions
P0, P1a--P1b, P2a--P2b, and P3a--P3d identified in Fig.~\ref{stab1}.
The results are shown for $\Delta_1=\Delta_2=1$, $g_1=g_2=0.1,0.5$, and $\Omega=1,2$.
The uncoupled-oscillator configuration, $\alpha=\beta=0$, is included in
each panel as a reference. The interaction parameters used for each dynamical region 
are specified as follows:
Top left four panels correspond to P1a and P1b; \ie, complex eigenvalues with dominant
real part and for the particular cases (a): $\alpha = -\sqrt{\Omega} - 0.5$ and 
$\beta = \sqrt{\Omega} + 0.5$ and (b): $\alpha = \sqrt{\Omega} + 0.5$, and 
$\beta = -\sqrt{\Omega} - 0.5$. The top right four panels corresponds to 
P2a and P2b; \ie, purely imaginary eigenvalues and for the two particular cases (a): 
$(\alpha, \beta) = (\sqrt{\Omega} + 0.5, \sqrt{\Omega} + 0.5)$,  
and (b): $(\alpha, \beta) = (-\sqrt{\Omega}-0.5, -\sqrt{\Omega}-0.5)$. The 
bottom panels correspond to the cases P3a and P3b (bottom left panels) and 
P3c and P3d (bottom right panels); \ie, mixed eigenvalues, two purely imaginary and two 
complex with dominant real part, and for the four particular cases (a): $\alpha = 0$ and 
$\beta = \sqrt{\Omega} + 0.5$, (b): $\alpha = -\sqrt{\Omega} - 0.5$ and $\beta = 0$, 
(c): $\alpha = 0$ and $\beta = -\sqrt{\Omega} - 0.5$, and (d): $\alpha = \sqrt{\Omega} + 0.5$, 
$\beta = 0$.}
\end{figure*}

To investigate the influence of these regimes on the two-qubit entanglement,
we evaluate the concurrence $\mc{C}(t)$ for representative values
of $\alpha$ and $\beta$ belonging to each region. Figure~\ref{fig2}
presents the corresponding dynamics for $\Delta_1=\Delta_2=1$,
$g_1=g_2=0.1,0.5$, and $\Omega=1,2$. In all cases, the uncoupled-oscillator 
configuration $\alpha=\beta=0$ is included as a reference.

The results reveal a strong dependence of the two-qubit entanglement dynamics
on the collective behavior of the oscillator subsystem. In the dynamically
unstable regions P1a and P1b, the concurrence exhibits a pronounced decay,
reflecting the rapid suppression of coherence between the initially entangled qubit states. 
As the qubit--oscillator coupling strengths increase, this decay becomes faster, with the 
profiles suggesting a crossover between approximately Gaussian and exponential-like
behavior over the time intervals considered. Similar behavior is observed in certain 
configurations of the mixed dynamical regions P3.

A qualitatively different behavior emerges in the oscillatory regions P2a and P2b. 
In these configurations, the concurrence exhibits persistent oscillations, reflecting the
recurrent exchange of quantum correlations between the qubit and oscillator degrees of 
freedom. Of particular interest is the region P2b, where the oscillation amplitude is 
substantially reduced compared with the uncoupled-oscillator case. Since the concurrence 
remains closer to its initial value, this behavior indicates a suppression of entanglement 
degradation induced by the oscillator--oscillator interaction.

This observation suggests that the collective oscillator dynamics can effectively reduce 
the dephasing experienced by the two-qubit subsystem, despite the continued presence of
the qubit--oscillator interactions. In this sense, suitable oscillator couplings provide 
a mechanism for modulating and protecting two-qubit entanglement through the internal
dynamics of auxiliary bosonic modes.

The mixed dynamical regions P3 exhibit a more varied behavior, reflecting the coexistence 
of oscillatory and unstable contributions to the oscillator dynamics. In particular,
the configurations P3b and P3c display oscillatory concurrence for certain parameter choices,
whereas a predominantly decaying behavior emerges as $\Omega$ is increased for the 
representative couplings considered. Conversely, the regions P3a and P3d exhibit
a pronounced decay of concurrence in the cases shown.

These results indicate that the entanglement dynamics is not determined solely by whether 
the oscillators are coupled, but also by the dynamical regime established through the
interaction parameters $\alpha$ and $\beta$. The distinction between oscillatory and unstable 
behavior provides a useful criterion for identifying configurations with qualitatively 
different entanglement evolution. Nevertheless, dynamical stability alone does not guarantee
entanglement protection, which depends on the particular choice of interaction parameters and
the resulting qubit--oscillator dynamics.

\subsubsection{\label{sec:markovian-transition}
Entanglement revivals and non-Markovian behavior}

The oscillatory behavior of concurrence observed in several dynamical regimes indicates that
entanglement initially degraded by the interaction with the oscillator subsystem can 
subsequently be recovered. As discussed in Ref.~\cite{Lopez26}, the dynamics of a
related qubit--oscillator configuration exhibits a redistribution of quantum correlations 
among different partitions of the composite system. The present results suggest that this 
redistribution can be modified through the interactions between the oscillators, thereby 
controlling the occurrence and amplitude of entanglement revivals.

An important consequence of such revivals is their connection with non-Markovian
effects in the reduced two-qubit dynamics. In particular, for the dephasing model
considered here, the reduced evolution within the subspace spanned by
$|g_1e_2\rangle$ and $|e_1g_2\rangle$ is governed by the coherence function
$f(t)$. The trace distance between the two initially orthogonal Bell states
$|\Psi^\pm\rangle$ evolves as
\begin{equation}
D\bigl(\varrho_+(t),\varrho_-(t)\bigr)
=|f(t)|,
\end{equation}
where $D(\varrho_1,\varrho_2) =\frac{1}{2}\mathrm{Tr}
|\varrho_1-\varrho_2|$.
Therefore, an increase in $|f(t)|$ corresponds to a temporary recovery of 
distinguishability between these states and provides a witness of information backflow 
according to the trace-distance criterion for non-Markovian dynamics
\cite{Breuer2009,Breuer2016}.

Since the concurrence of the initial Bell state satisfies $\mc{C}(t)=|f(t)|$, 
its revivals are directly associated with intervals of increasing trace
distance. Thus, the oscillatory concurrence regimes identified above exhibit 
signatures of non-Markovian reduced dynamics.
Conversely, configurations displaying monotonic concurrence decay do not exhibit 
this particular information-backflow signature, although such decay alone does not
exclude all possible non-Markovian effects.

From this perspective, tuning the oscillator interaction parameters allows us to modify 
the qualitative behavior of the reduced dynamics, ranging from regimes with pronounced
entanglement revivals to regimes dominated by irreversible-looking decay over the time 
intervals studied.
The latter are particularly evident when the oscillator dynamics contains unstable 
contributions, which can produce a rapid suppression of the qubit coherence.

\begin{figure}[ht]
 \centering
  \includegraphics[width=\linewidth]{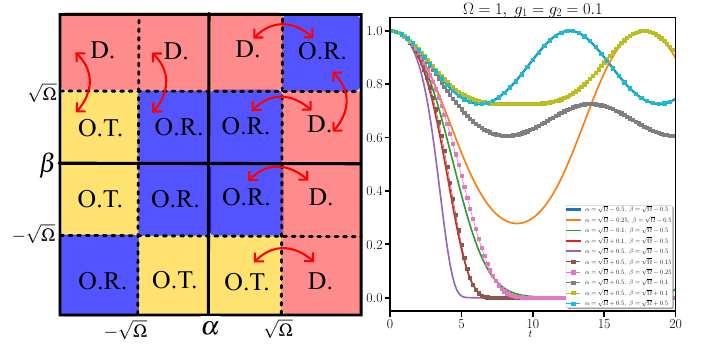}
\caption{\label{fig3} 
Control of two-qubit entanglement dynamics through the oscillator interaction 
parameters. The left panel classifies the observed concurrence behavior in the
$(\alpha,\beta)$ parameter space: O.R. denotes oscillatory dynamics with entanglement 
revivals, D. denotes predominantly decaying concurrence, and O.T. identifies 
configurations exhibiting oscillatory or decaying behavior depending on $\Omega$.
The right panel illustrates the change in concurrence dynamics as the interaction 
parameters are varied between regions with entanglement revivals and regions 
dominated by concurrence decay.}
\end{figure}

Figure~\ref{fig3} illustrates this control mechanism. The left panel summarizes the 
different concurrence behaviors observed across the $(\alpha,\beta)$ parameter space.
The regions labeled O.R. correspond to oscillatory concurrence with repeated revivals, 
whereas D. identifies configurations exhibiting predominantly decaying concurrence.
The regions labeled O.T. describe parameter configurations in which the observed 
behavior changes from oscillatory to decaying depending on the oscillator frequency 
$\Omega$.

The right panel of Fig.~\ref{fig3} shows how the concurrence dynamics changes as the 
oscillator coupling parameters are varied across these regions. The evolution passes 
from an oscillatory regime exhibiting entanglement revivals to a regime dominated by 
concurrence decay, and subsequently to another oscillatory regime. This illustrates
how the interaction between the auxiliary oscillators can be used to control the 
occurrence of entanglement revivals and the associated signatures of non-Markovian 
information backflow.

These results establish a connection between the dynamical properties of the auxiliary 
oscillator subsystem and the temporal behavior of quantum correlations in the two-qubit 
subsystem. In particular, the possibility of suppressing entanglement degradation through 
suitable oscillator interactions motivates investigating whether this control mechanism 
remains effective when environmental dissipation is included.

%%%%%%%%%%%%%%%%%%%%%%%%%%%%%%%%%%%%%%%%%%%%%%%%%%%%%%%%%%%%%%%%%%%%%%%%%

\subsection{\label{sec:local-dissipation}
Entanglement protection in the presence of
oscillator dissipation}

The results presented in the preceding section demonstrate that the interaction between the
oscillators can substantially modify the entanglement dynamics of the two-qubit subsystem.
In particular, certain oscillatory regimes exhibit a reduced degradation of concurrence
compared with the uncoupled-oscillator configuration, suggesting an effective
suppression of the dephasing induced by the auxiliary bosonic degrees of freedom.

An important question is whether this protective behavior persists when the oscillators 
interact with external dissipative environments. Since the environmental coupling acts 
directly on the oscillator subsystem, the qubits experience its influence indirectly
through their dephasing interactions. Consequently, modifying the collective oscillator 
dynamics may also influence the way environmental noise affects the two-qubit entanglement.

To investigate this possibility, we consider the local and global dissipative
descriptions introduced in Sec.~\ref{model}. The former accounts for dissipation 
through independent oscillator reservoirs, while the latter describes dissipation in terms 
of the collective normal modes of the interacting oscillator subsystem.
In both cases, we examine whether suitable values of the tunable interaction parameters 
$\alpha$ and $\beta$ can enhance the preservation of two-qubit entanglement.

\subsubsection{Entanglement dynamics under
local dissipation}

We first investigate the influence of local thermal dissipation on the two-qubit entanglement 
dynamics. The analytical solutions of the corresponding master equation, expressed in the 
characteristic-function representation, are presented in Appendix~\ref{app2}. Here, we focus
on the resulting evolution of the two-qubit concurrence.

\begin{figure}[ht]
 \centering
\includegraphics[width=\linewidth]{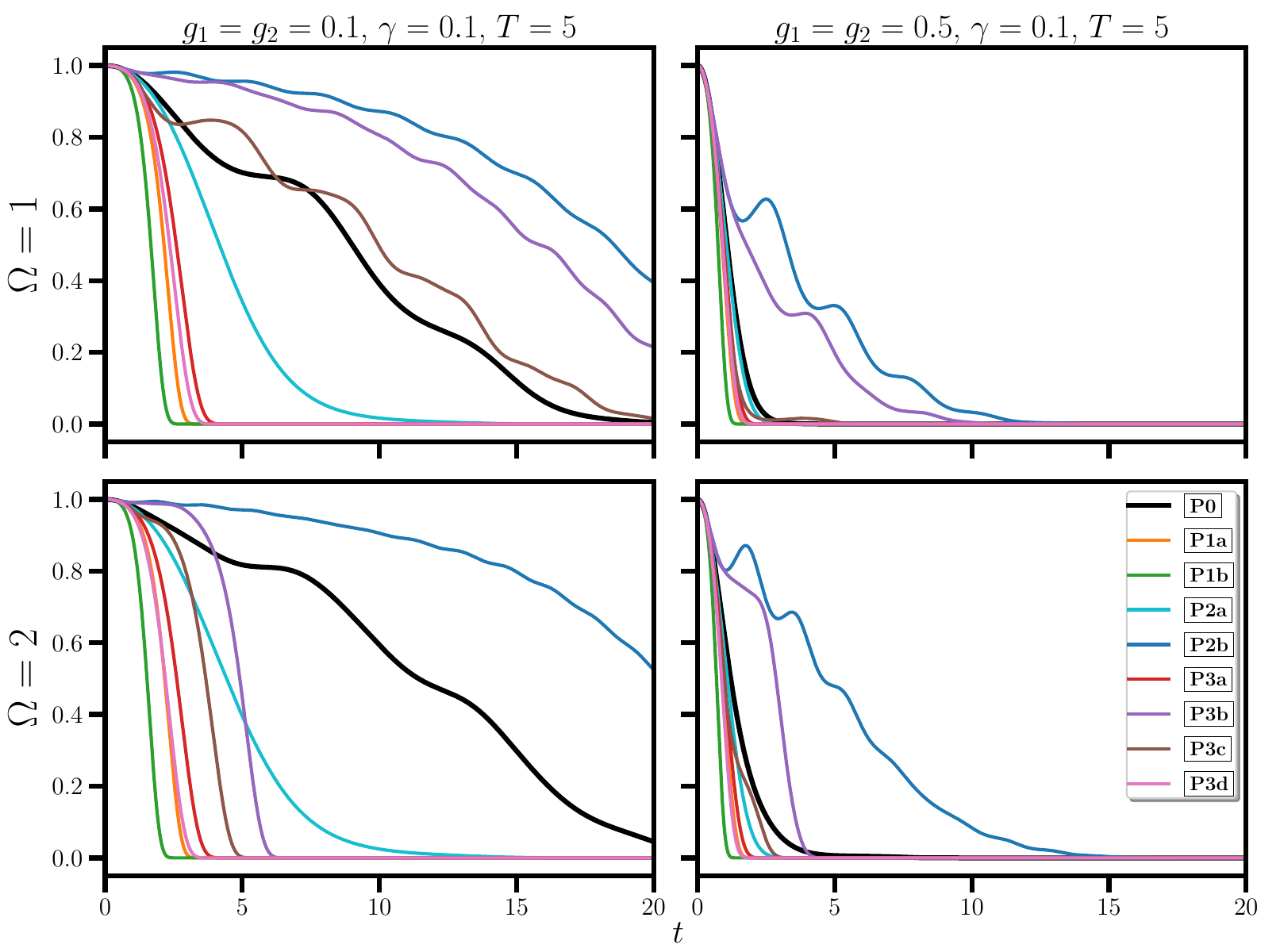}
  \caption{\label{fig4}  Time evolution of the two-qubit concurrence
$\mc{C}(t)$ in the presence of local thermal dissipation. The panels correspond to different 
oscillator frequency ratios $\Omega=1,2$ and qubit--oscillator coupling strengths 
$g_1=g_2=0.1,0.5$. The black curve represents the uncoupled-oscillator configuration
P0, $(\alpha,\beta)=(0,0)$. The remaining curves correspond to the representative interaction 
parameters P1a: $(-\sqrt{\Omega}-0.5,\sqrt{\Omega}+0.5)$; P1b: $(\sqrt{\Omega}+0.5,-\sqrt{\Omega}-0.5)$;
P2a: $(\sqrt{\Omega}+0.5,\sqrt{\Omega}+0.5)$; P2b: $(-\sqrt{\Omega}-0.5,-\sqrt{\Omega}-0.5)$;
P3a: $(0,\sqrt{\Omega}+0.5)$; P3b: $(-\sqrt{\Omega}-0.5,0)$; P3c: $(0,-\sqrt{\Omega}-0.5)$;
and P3d: $(\sqrt{\Omega}+0.5,0)$. The remaining parameters are $\Delta_1=\Delta_2=1$,
$\gamma_a=\gamma_b=0.1$, and $T_a=T_b=5$.}
\end{figure}

Figure~\ref{fig4} shows the concurrence $\mc{C}(t)$ for representative oscillator interaction 
parameters belonging to the different dynamical regions identified in the preceding section. 
The results are presented for $\Omega=1,2$ and $g_1=g_2=0.1,0.5$, under the local dissipative 
conditions specified in the figure.

In each panel, the thicker black curve corresponds to the reference configuration 
$\alpha=\beta=0$, in which the oscillators do not interact. This provides a direct
comparison with the entanglement dynamics obtained when the internal oscillator interactions 
are introduced, while keeping the remaining parameters fixed.

The results reveal that the influence of local dissipation on the two-qubit entanglement 
depends strongly on the dynamical regime established by the oscillator interaction parameters.
In several configurations, the concurrence decreases more rapidly than in the 
uncoupled-oscillator case, indicating an enhancement of entanglement degradation. In contrast,
other configurations exhibit a substantial suppression of concurrence decay, demonstrating 
that oscillator interactions can mitigate the indirect influence of environmental dissipation 
on the qubit subsystem.

Of particular interest is the oscillatory region P2b, which exhibits a pronounced 
preservation of two-qubit concurrence over the time intervals considered. Compared with 
the uncoupled-oscillator configuration, the entanglement remains substantially higher, even 
though the oscillator subsystem is subject to thermal dissipation.
This behavior is observed for the representative parameters investigated, suggesting that 
the protective effect identified in the nondissipative case can persist in the presence
of local environmental noise.

Other dynamical regions also exhibit partial enhancement of the two-qubit entanglement, 
although the effect is generally less pronounced or more dependent on the remaining 
system parameters. These observations indicate that entanglement protection is not
a generic consequence of introducing an oscillator--oscillator interaction, but rather 
depends on the specific collective dynamics established by the parameters $\alpha$ and
$\beta$.

The results are consistent with an effective suppression of the dephasing experienced by 
the qubits through the collective oscillator dynamics. Although the oscillators remain 
coupled to their respective thermal environments, their mutual interaction modifies the 
evolution of the qubit coherence, allowing the concurrence to remain higher than in the 
absence of oscillator--oscillator coupling.

We next examine whether a similar behavior occurs when dissipation is formulated in terms 
of the collective normal modes of the interacting oscillator subsystem.

\subsubsection{\label{sec:global-dissipation}
Entanglement dynamics under global dissipation}

We now investigate the robustness of two-qubit entanglement under a global dissipative 
description, in which the environmental interaction is formulated in terms of the 
collective normal modes of the coupled-oscillator subsystem. The corresponding master 
equation, together with its characteristic-function representation, is derived in 
Appendix~\ref{app3}.

Unlike the local dissipative model, the global description explicitly incorporates the
oscillator--oscillator interaction into the definition of the dissipative operators. 
This approach is particularly relevant when the normal-mode frequencies are sufficiently
separated for the secular approximation to suppress cross-frequency dissipative terms.

In the following, we restrict our analysis to the positive-energy stability region,
$|\alpha|<1$ and $|\beta|<1$, for the resonant oscillator configuration $\Omega=1$. 
In this region, the quadratic oscillator Hamiltonian is bounded from below and 
admits a decomposition into positive-energy normal modes.
We focus on the symmetric interaction $\alpha=\beta=\eta$, for which the oscillator
coupling reduces to an excitation-exchange interaction. The normal-mode frequencies
take the particularly simple form
\begin{equation}
\tilde{\omega}_1=1-\eta,\qquad \tilde{\omega}_2=1+\eta.
\end{equation}

Of particular interest are configurations approaching the stability boundaries
$\eta=\pm1$, without crossing them. We consider $\eta=-0.99$ and $\eta=0.99$,
which approach the dynamically oscillatory regions P2b and P2a, respectively, 
while remaining within the positive-energy region P0.

\begin{figure}[ht]
 \centering
\includegraphics[width=\linewidth]{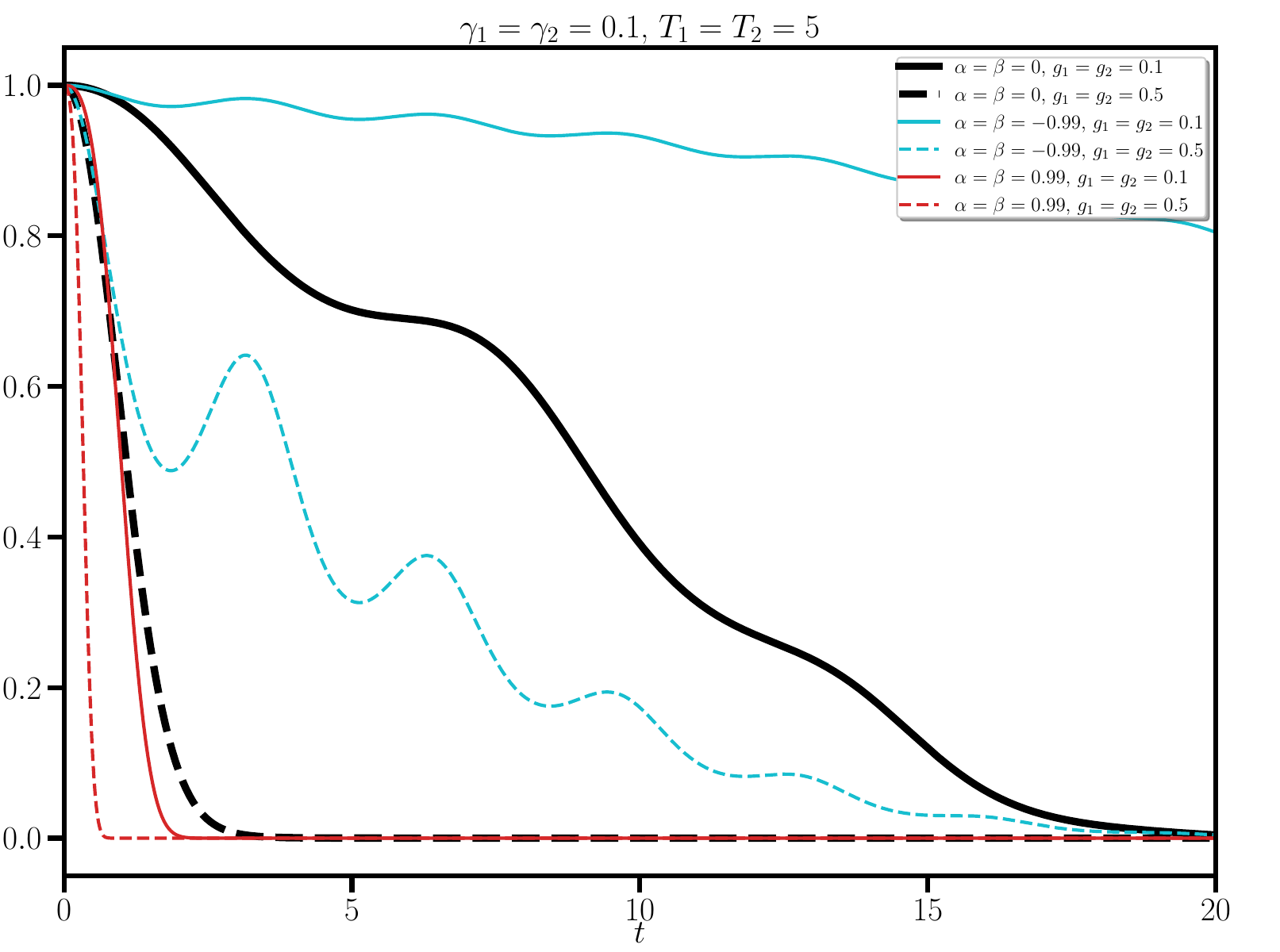}
\caption{\label{fig5}
Time evolution of the two-qubit concurrence $\mc{C}(t)$ in the presence of oscillator
dissipation, for resonant oscillators $\Omega=1$. The thick black curves
correspond to the uncoupled-oscillator configuration, $\alpha=\beta=0$,
described through the local dissipative model. The cyan and red curves correspond
to interacting oscillators with $\alpha=\beta=-0.99$ and $\alpha=\beta=0.99$, 
respectively, described through the global normal-mode master equation. Solid and 
dashed curves represent qubit--oscillator coupling strengths $g_1=g_2=0.1$
and $g_1=g_2=0.5$, respectively. The dissipative parameters are $\gamma_1=\gamma_2=0.1$ 
and $T_1=T_2=5$, while the qubit frequencies are fixed at $\Delta_1=\Delta_2=1$.
Both interacting configurations lie within the positive-energy stability region, 
approaching the negative and positive interaction boundaries, respectively.
}
\end{figure}

Figure~\ref{fig5} presents the time evolution of the two-qubit concurrence for 
these configurations, together with the uncoupled-oscillator reference
$\alpha=\beta=0$. Two different values of the qubit--oscillator coupling 
strength are considered, $g_1=g_2=0.1$ and $g_1=g_2=0.5$, while the dissipative
parameters are fixed at $\gamma_1=\gamma_2=0.1$ and $T_1=T_2=5$.

The uncoupled-oscillator reference is described using the local dissipative
model, whereas the interacting configurations are treated using the global 
normal-mode master equation. This distinction is important because, at 
$\alpha=\beta=0$, the normal-mode frequencies become degenerate and a 
microscopic common-bath description generally requires retaining collective 
dissipative contributions. Consequently, the reference curves provide
a comparison between the corresponding dissipative scenarios rather than the
strict zero-coupling limit of a single common-reservoir model.

As shown in Fig.~\ref{fig5}, the concurrence exhibits a pronounced dependence on 
both the oscillator interaction parameters and the qubit--oscillator coupling strength.
For $\eta=-0.99$, a substantial enhancement of entanglement preservation is observed
relative to the uncoupled reference, particularly for $g_1=g_2=0.1$.
In this configuration, the concurrence remains close to its initial value,
displaying only relatively small oscillations and gradual degradation over the time 
interval considered.

For the stronger qubit--oscillator coupling, $g_1=g_2=0.5$, the same oscillator configuration 
still exhibits pronounced entanglement revivals, although the overall concurrence
decreases substantially faster. This indicates that the effectiveness of the protective 
mechanism depends not only on the collective oscillator dynamics but also on the strength
of the coupling between the qubits and the auxiliary oscillator subsystem.

In contrast, the configuration $\eta=0.99$ exhibits a rapid suppression of concurrence for 
both values of the qubit--oscillator coupling. Thus, approaching opposite boundaries
of the stable interaction region leads to qualitatively different entanglement dynamics, 
despite the symmetric form of the oscillator--oscillator interaction.

This behavior can be understood by examining the normal modes responsible
for the dephasing of the initially entangled qubits. For equal qubit--oscillator couplings,
$g_1=g_2=g$, the interaction Hamiltonian can be expressed in the normal-mode basis as
\begin{equation}
\begin{aligned}
H_{\mathrm{coupling}}
={}&
\frac{g}{\sqrt{2}} \left(\sigma_z^{(2)} -\sigma_z^{(1)}\right)X_1
 + \frac{g}{\sqrt{2}} \left(\sigma_z^{(1)} + \sigma_z^{(2)}\right)X_2.
\end{aligned}
\end{equation}

For the initially prepared Bell state
$|\Psi^+\rangle = (|e_1g_2\rangle+|g_1e_2\rangle)/\sqrt{2}$,
the operator $\sigma_z^{(1)}+\sigma_z^{(2)}$ vanishes within the dynamically
accessible two-qubit subspace. Consequently, only the antisymmetric normal mode 
$X_1$ contributes directly to the dephasing of the corresponding two-qubit coherence.

For $\eta=-0.99$, the antisymmetric mode has frequency $\tilde{\omega}_1=1.99$,
while the symmetric mode becomes soft, with $\tilde{\omega}_2=0.01$.
The opposite configuration occurs for $\eta=0.99$, where the antisymmetric mode 
becomes soft and the symmetric mode acquires the higher frequency.

Since the Bell-state coherence couples exclusively to the antisymmetric mode, 
the effective strength of the conditional oscillator displacement is controlled by the
ratio $g/\tilde{\omega}_1$. Increasing the frequency of this mode therefore suppresses 
the relative influence of the qubit--oscillator interaction, whereas reducing its frequency
enhances the resulting dephasing.

This provides a physical explanation for the enhanced concurrence observed near the 
negative interaction boundary and its rapid degradation near the positive boundary. 
Furthermore, the pronounced difference between the curves corresponding to $g=0.1$ 
and $g=0.5$ highlights the importance of the relative strength of the qubit--oscillator
coupling and the relevant normal-mode frequency.

The results suggest that entanglement protection can be achieved by engineering
the collective oscillator dynamics so that the mode coupled to the relevant qubit 
coherence becomes effectively less susceptible to qubit-induced conditional 
displacement. In the symmetric configuration considered here, this occurs while
the complementary soft mode is decoupled from the Bell-state coherence.

We emphasize that the global dissipative description is based on the oscillator-normal-mode
approximation discussed in Appendix~\ref{app3}. In particular, the proximity of one normal-mode
frequency to zero requires care in assessing the weak-coupling and secular approximations.
The bath-induced Lamb shift, which produces perturbative renormalizations of the quadratic
oscillator Hamiltonian, has been neglected in the numerical results.

Nevertheless, within the adopted dissipative description, the results demonstrate that 
suitable oscillator--oscillator interactions can substantially suppress the degradation 
of two-qubit entanglement. The protective effect is therefore not restricted to idealized
nondissipative dynamics and can persist in the presence of environmental noise acting on
the auxiliary bosonic modes.

\section{\label{conclusion}Summary and conclusions}

In this work, we have investigated the modulation and protection of entanglement in a two-qubit 
system coupled through dephasing interactions to two mutually interacting harmonic oscillators.
By introducing independently tunable position--position and momentum--momentum couplings between 
the oscillators, we have explored how their collective dynamics influences the evolution of quantum 
correlations in the two-qubit subsystem.

Using the characteristic-function representation, we obtained analytical solutions for the dynamics 
of the composite system, both in the absence and in the presence of oscillator dissipation. 
This approach allows the two-qubit concurrence to be determined directly from the evolution
of a single coherence function, providing a convenient framework for investigating the influence
of the oscillator interactions.

Our results establish a connection between the dynamical properties of the coupled oscillators 
and the entanglement evolution of the two-qubit subsystem. Depending on the oscillator
interaction parameters, the concurrence exhibits qualitatively different behaviors, including
pronounced decay, oscillations, and entanglement revivals. The latter are also associated
with signatures of non-Markovian information backflow in the reduced two-qubit dynamics.

Of particular interest is the identification of interaction configurations that substantially
suppress the degradation of two-qubit entanglement. We have shown that this protective behavior 
can occur within the positive-energy stability region of the oscillator Hamiltonian, without 
requiring dynamically unstable or negative-energy oscillator configurations.

The physical origin of this effect can be understood through the collective normal-mode 
structure of the oscillator subsystem. For symmetric oscillator interactions and equal
qubit--oscillator coupling strengths, the initially entangled Bell-state coherence
couples exclusively to the antisymmetric oscillator mode. By tuning the oscillator
interaction parameters, the frequency of this mode can be increased, reducing the
relative influence of the qubit-induced conditional displacements and thereby
suppressing dephasing.

In particular, configurations approaching the negative interaction boundary exhibit
enhanced preservation of concurrence, whereas approaching the opposite boundary results
in stronger entanglement degradation. The effectiveness of this mechanism also depends
on the qubit--oscillator coupling strength, with weaker coupling generally favoring 
greater preservation of entanglement in the configurations considered.

We further investigated the robustness of this behavior under environmental dissipation
acting directly on the oscillator degrees of freedom. Both local and global dissipative 
descriptions were considered, with the global model formulated in terms of the normal modes
of the interacting oscillators. The results show that suitable oscillator interactions can
maintain substantial two-qubit entanglement under the adopted dissipative conditions.

The global analysis also highlights the importance of the spectral structure of the 
auxiliary oscillator system in determining the applicability of the dissipative 
approximation. In particular, the separation of normal-mode frequencies and the positivity 
of the quadratic Hamiltonian provide important conditions for the physical interpretation
of the results.

Overall, our findings suggest that engineering interactions between auxiliary bosonic modes
offers a promising strategy for controlling and protecting entanglement in discrete-variable
quantum systems. The connection between normal-mode dynamics and the suppression of qubit 
dephasing provides a physical mechanism that may be relevant to the design of controllable 
hybrid quantum architectures.

Future investigations could address the robustness of this mechanism against asymmetric 
qubit--oscillator couplings, additional sources of environmental noise, and more general
oscillator--reservoir interactions.

\appendix
\section{\label{app1} Stability analysis of the two-oscillator system}
The dynamics of the two-interacting oscillator system in the absence of dissipation, 
whose Hamiltonian is given by: 
\begin{equation}
  H_{\mathrm{osc's}} = 
H_{\mathrm{oscillators}} + H_{\mathrm{oscillator\_interaction}},
\end{equation}
where $H_{\mathrm{oscillators}}$ and $H_{\mathrm{oscillator\_interaction}}$ are 
given in Eqs. (\ref{hoscillators}) and (\ref{hinteractionoscillators}) 
posses a rich structure of dynamical regimes. 
The dynamics of the two-oscillator system can be 
characterized through the stability analysis of the system,
which in the Fourier phase-space context, \ie, the characteristic function representation, 
it can be completely determined solely through 
the stability matrix (\ref{disstabilityMat}) when $\gamma_1=\gamma_2=0$; \ie:
\begin{equation}\label{stabilityMat}
{\bf A} = \left(
  \begin{array}{cccc}
     0 & 1 & 0 & \alpha\\
     -1 & 0 & -\beta & 0\\
     0 & \alpha & 0 & \Omega\\
     -\beta & 0 & -\Omega & 0
   \end{array}\right)\,.
\end{equation}
It is worth mentioning that the dynamics of the two-oscillator system in the 
Fourier phase-space representation is connected to the traditional quantum phase-space representation, 
\ie, the Wigner function representation of the system, through the following relations among the
stability matrix and transition matrix of the system: ${\bf A}_c = -{\bf A}^{T}$ and 
$\Phi_c(t) = \Phi^{T}(-t)$, where ${\bf A}_c$ and $\Phi_c(t)$ 
are the stability and transition matrices that govern the dynamics of the 
first moments (position and momentum) of the two-oscillator system in the Wigner function 
representation. 

The eigenvalues of the stability matrix (\ref{stabilityMat}) are given by:
\begin{eqnarray}\label{eigenvalues}
\lambda_{1,2} &=& \pm { 1\over \sqrt{2}} \sqrt{-\Omega^2 - 2\alpha\beta - 1 - r}\\
\lambda_{3,4} &=& \pm { 1\over \sqrt{2}} \sqrt{-\Omega^2 - 2\alpha\beta - 1 + r},
\end{eqnarray}
where 
\begin{equation}
r = \sqrt{(\Omega^2 + 2\alpha\beta + 1)^2 - 4(\alpha^2-\Omega)(\beta^2-\Omega)};
\end{equation}
and they can be purely real (instability and exponential divergence), 
purely imaginary (oscillatory behavior), or mixed (two real and two imaginary), 
depending on the values of the tunable coupling parameters $\alpha$ and $\beta$.

Moreover, the determinant of the stability matrix (\ref{stabilityMat})
\begin{equation}
\det({\bf A}) = (\alpha^2-\Omega)(\beta^2-\Omega),
\end{equation}
shows that the system undergoes a double-zero bifurcation~\cite{} when
$\alpha = \pm \sqrt{\Omega}$ or $\beta = \pm \sqrt{\Omega}$ or both, 
defining a bifurcation threshold from which the system transitions 
from stable to unstable dynamics.

\begin{figure*}[ht]
\centering
  \includegraphics[width=\textwidth]{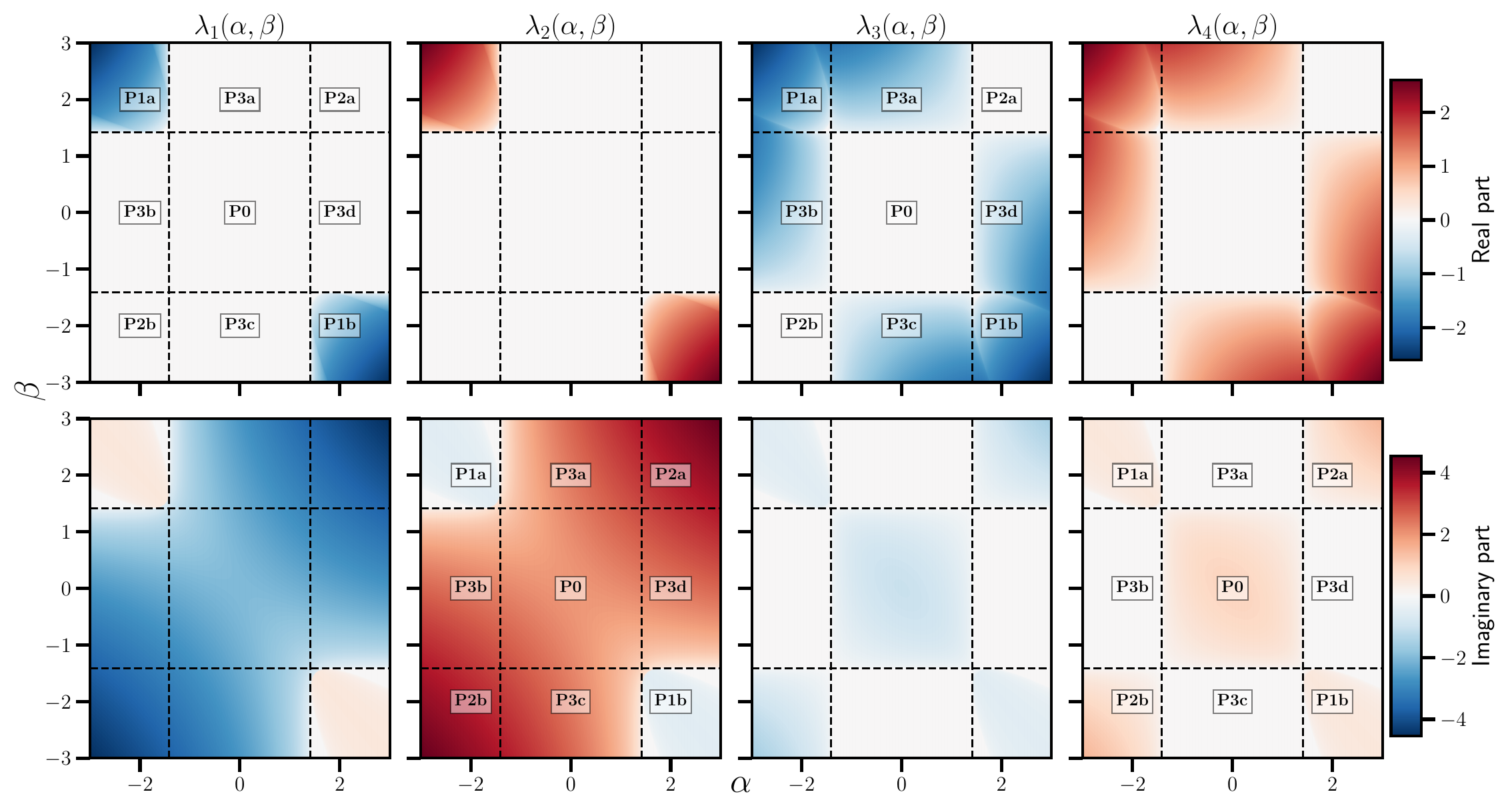}
  \caption{\label{stab1}
  Real and imaginary parts of the four eigenvalues of the stability matrix (\ref{stabilityMat}) as a function of the tunable 
  coupling parameters $\alpha$ and $\beta$, for a fixed value of $\Omega = 2$.}
\end{figure*}

In figure \ref{stab1} we have depicted the real and imaginary parts of the 
four eigenvalues of the stability matrix (\ref{stabilityMat}) as 
a function of the tunable coupling parameters $\alpha$ and $\beta$, 
for a fixed value of $\Omega = 1$. 
In it, one can observe 
regions of stability (where purely imaginary eigenvalues are present) and 
instabilities (where purely real or complex eigenvalues are present), 
as well as the transition limits from imaginary to complex (real and imaginary parts) 
eigenvalues, which correspond to the bifurcation thresholds at 
$\alpha = \pm \sqrt{\Omega}$ and $\beta = \pm \sqrt{\Omega}$. Particularly, we 
have in addition depicted in the figure specific cases which are 
addressed in the main text: 
\begin{itemize}
\item [\fbox{{\bf P0}} :] $\alpha = 0$ and $\beta = 0$ (no interaction between the oscillators),
 and $ -\sqrt{\Omega} < \alpha,\beta < \sqrt{\Omega}$ (purely imaginary eigenvalues)
\item [\fbox{{\bf P1}} :] $\alpha < -\sqrt{\Omega}$ and $\beta > \sqrt{\Omega}$ or
$\alpha > \sqrt{\Omega}$ and $\beta < -\sqrt{\Omega}$ (purely real or complex eigenvalues)
\item [\fbox{{\bf P2}} :] $\alpha,\beta > \sqrt{\Omega}$ or $\alpha \beta < -\sqrt{\Omega}$  
(purely imaginary eigenvalues) 
\item [\fbox{{\bf P3}} :] $\alpha > \sqrt{\Omega}$ and $ -\sqrt{\Omega}< \beta < \sqrt{\Omega}$ 
or $ -\sqrt{\Omega} < \alpha < \sqrt{\Omega}$ and $ \beta > \sqrt{\Omega}$ 
or  $\alpha < -\sqrt{\Omega}$ and $ -\sqrt{\Omega} < \beta < \sqrt{\Omega}$  
or $ -\sqrt{\Omega} < \alpha < \sqrt{\Omega}$ and $ \beta < -\sqrt{\Omega}$ 
(two eigenvalues are purely imaginary and two are purely real) 
\end{itemize}
These regions of stability and instability in the two-oscillator system have a direct 
impact on the dynamics of entanglement in the two-qubit subsystem, 
as it has been shown in the main body of the text.

\section{\label{app2} Analytical solutions}

We follow the same lines as the derivation of the solutions given in \cite{Lopez26}, but with the inclusion 
of the coupling between the two oscillators together with the dissipative configurations treated in this work. 
The form of the procedure and the structure of the solutions, is very similar as in \cite{Lopez26}, 
but only the transition matrix and additional Gaussian elements corresponding to the 
diffusion processes in the characteristic function is what changes.

The projection of the master equation (\ref{master_eq}) of the system into the two-qubit computational
basis states as described in (\ref{densmatrix_elements}) yields the following set of differential 
equations:
\begin{flalign}\label{rho11}
  &\rmi \dot{\varrho}_{11} = [H_{\mathrm{osc's}},\varrho_{11}] - g_1[\h{x}_1,\varrho_{11}]  
  - g_2[\h{x}_2,\varrho_{11}] + \rmi \mathcal{D}_{11}[\varrho],&\\\nonumber
  &\rmi \dot{\varrho}_{12} =-\Delta_2 \varrho_{12} + [H_{\mathrm{osc's}},\varrho_{12}]
  - g_1[\h{x}_1,\varrho_{12}]&  \\
  & \hspace{4.2cm}- g_2\{\h{x}_2,\varrho_{12}\}+ \rmi \mathcal{D}_{12}[\varrho],&\\\nonumber
  &\rmi \dot{\varrho}_{13} =-\Delta_1 \varrho_{13} + [H_{\mathrm{osc's}},\varrho_{13}]
  - g_1\{\h{x}_1,\varrho_{13}\}  &\\
  & \hspace{4.3cm}-  g_2[\h{x}_2,\varrho_{13}] + \rmi \mathcal{D}_{13}[\varrho],&\\\nonumber
  &\rmi \dot{\varrho}_{14} =-(\Delta_1+\Delta_2) \varrho_{14} + [H_{\mathrm{osc's}},\varrho_{14}]&\\
  &\hspace{2.2cm}- g_1\{\h{x}_1,\varrho_{14}\}  - g_2\{\h{x}_2,\varrho_{14}\} + \rmi \mathcal{D}_{14}[\varrho],&
\end{flalign}

\begin{flalign}\label{rho22}
  &\rmi \dot{\varrho}_{22} = [H_{\mathrm{osc's}},\varrho_{22}]
  - g_1[\h{x}_1,\varrho_{22}]  + g_2[\h{x}_2,\varrho_{22}]  + \rmi \mathcal{D}_{22}[\varrho],&\\\nonumber
  &\rmi \dot{\varrho}_{23} =-(\Delta_1-\Delta_2) \varrho_{23} + [H_{\mathrm{osc's}},\varrho_{23}]&\\
  &\hspace{2.2cm}- g_1\{\h{x}_1,\varrho_{23}\}  + g_2\{\h{x}_2,\varrho_{23}\} + \rmi \mathcal{D}_{23}[\varrho],&\\\nonumber
  &\rmi \dot{\varrho}_{24} =-\Delta_1\varrho_{24} + [H_{\mathrm{osc's}},\varrho_{24}]
  - g_1\{\h{x}_1,\varrho_{24}\} & \\
  &\hspace{4.3cm}+ g_2[\h{x}_2,\varrho_{24}] + \rmi \mathcal{D}_{24}[\varrho],&
\end{flalign}

\begin{flalign}\label{rho33}
  &\rmi \dot{\varrho}_{33} = [H_{\mathrm{osc's}},\varrho_{33}]
  + g_1[\h{x}_1,\varrho_{33}]  -  g_2[\h{x}_2,\varrho_{33}] +  \rmi \mathcal{D}_{33}[\varrho],&\\\nonumber
  &\rmi \dot{\varrho}_{34} = - \Delta_2 \varrho_{34} + [H_{\mathrm{osc's}},\varrho_{34}]
  + g_1[\h{x}_1,\varrho_{34}]  \\
  &\hspace{4.2cm}- g_2\{\h{x}_2,\varrho_{34}\} + \rmi \mathcal{D}_{34}[\varrho],&
\end{flalign}
\begin{flalign}\nonumber
  &\rmi \dot{\varrho}_{44} = [H_{\mathrm{osc's}},\varrho_{44}]
  + g_1[\h{x}_1,\varrho_{44}]  + g_2[\h{x}_2,\varrho_{44}] + \rmi \mathcal{D}_{44}[\varrho],&\\\label{rho44}.
\end{flalign}
In the non-dissipative case, the dynamics of the system is only described by the unitary part 
of the dynamics for which the dissipative part is absent, \ie, $\mathcal{D}_{ij}[\varrho] = 
\la i|\mathcal{D}[\varrho]|j\ra = 0$.
On the other hand, the distinct configurations of dissipation are encoded in the different forms of 
the dissipator $\mathcal{D}[\varrho]$ and the form it takes in its projections to the different matrix 
elements of the density operator.
Moving into the characteristic function frame can be done 
by following the following rules of transformation:
\begin{eqnarray}\label{xpr}
\h{x}^n \h{p}^m\, \varrho &\mapsto&
   \Big( \frac{s}{2} - \rmi\partial_k\Big)^{n}
   \Big( \frac{-k}{2} - \rmi\partial_s\Big)^{m}\! \!\!\!w(k,s),\\
\varrho\, \h{x}^{n}\h{p}^{m} &\mapsto&
   \Big( \frac{-s}{2} - \rmi \partial_k \Big)^n
   \Big( \frac{k}{2} - \rmi\partial_s \Big)^m\! \!\!\! w(k,s),\\
\h{x}^{n}\,\varrho\, \h{p}^{m} &\mapsto&
   \Big( \frac{s}{2} - \rmi \partial_k\Big)^{n}
   \Big( \frac{k}{2} - \rmi\partial_s\Big)^{m}\! \!\!\!w(k,s),\\\label{prx}
\h{p}^{m}\,\varrho\, \h{x}^{n} &\mapsto&
   \Big( \frac{-s}{2} - \rmi \partial_k\Big)^n
   \Big( \frac{-k}{2} - \rmi\partial_s \Big)^m\! \!\!\!w(k,s) ,
\end{eqnarray}
which yields the following set of 1st order partial differential equations:
\begin{flalign}
    &\mathcal{L}_o w_{11}(\vec{R},t) = \rmi( g_1 s_1 + g_2 s_2)w_{11}(\vec{R},t),&\\
    &(\mathcal{L}_o-2g_2\partial_{k_2})w_{12}(\vec{R},t) = \rmi (\Delta_2 + g_1s_1)w_{12}(\vec{R},t),& \\
    &(\mathcal{L}_o-2g_1\partial_{k_1})w_{13}(\vec{R},t) = \rmi (\Delta_1 + g_2s_2)w_{13}(\vec{R},t),& \\
    &(\mathcal{L}_o-2g_1\partial_{k_1}-2g_2\partial_{k_2})w_{14}(\vec{R},t) = \rmi (\Delta_1 + \Delta_2) w_{14}(\vec{R},t),& \\\nonumber
    &&\\
    &\mathcal{L}_o w_{22}(\vec{R},t) = \rmi( g_1 s_1 - g_2 s_2)w_{22}(\vec{R},t),&\\
    &(\mathcal{L}_o - 2g_1\partial_{k_1}+2g_2\partial_{k_2})w_{23}(\vec{R},t) = \rmi (\Delta_1 - \Delta_2)w_{23}(\vec{R},t), & \\
    &(\mathcal{L}_o - 2g_1\partial_{k_1})w_{24}(\vec{R},t) = \rmi (\Delta_1 - g_2s_2)w_{24}(\vec{R},t), & \\\nonumber
    &&\\
    &\mathcal{L}_o w_{33}(\vec{R},t) = -\rmi( g_1 s_1 - g_2 s_2)w_{33}(\vec{R},t),&\\
    &(\mathcal{L}_o - 2g_2\partial_{k_2}) w_{34}(\vec{R},t) = \rmi( \Delta_2 - g_1 s_1)w_{34}(\vec{R},t),&\\\nonumber
    &&\\
    &\mathcal{L}_o w_{44}(\vec{R},t) = -\rmi( g_1 s_1 + g_2 s_2)w_{44}(\vec{R},t),&
  \end{flalign}
where $\mathcal{L}_o$ is a linear differential operator whose form encodes the different dissipative 
configurations of the system. In the case of local dissipation this one takes the form: 
\begin{flalign}\nonumber
  &\mathcal{L}_o  = \partial_t + (s_1+\alpha s_2 + \gamma_1 k_1)\partial_{k_1} 
  - (k_1 + \beta k_2-\gamma_1 s_1)\partial_{s_1}& \\\nonumber
  & \hspace{1.15cm}+ (\Omega s_2 + \alpha s_1 + \gamma_2 k_2)\partial_{k_2} 
  - (\Omega k_2+ \beta k_1 - \gamma_2 s_2)\partial_{s_2}&\\\label{local_lo}
  &\hspace{1.15cm}-{1\over 2} \, \vec{R}^{T} \, \mathrm{M}_D\,  \vec{R}\,, &
\end{flalign}
where %$\mathrm{M}_D$
\begin{equation}
\mathrm{M}_D = {1\over 2}\left(
  \begin{array}{cccc}
     \gamma_1 N_1 & 0 & 0 & 0\\
     0 & \gamma_1 N_1 & 0 & 0\\
     0 & 0 & \gamma_2 N_2 & 0\\
     0 & 0 & 0 & \gamma_2 N_2
   \end{array}\right)\,,
\end{equation}
is the diffusion matrix of the local non-correlated dissipative environment. 
For the global correlated dissipative environment, we will be focusing on the 
resonant case where $\Omega = 1$, although non-resonant cases could also be considered 
(see Appendix \ref{app3} for reference). 
In this case the linear differential operator takes the following form:
\begin{equation}\label{global_lo}
  \begin{aligned}
  \mathcal{L}_o  ={}&
  \partial_t + (s_1+\alpha s_2 + \gamma_+ k_1 + \gamma_- k_2)\partial_{k_1}\\
  & - (k_1 + \beta k_2 - \gamma_+ s_1 - \gamma_- s_2)\partial_{s_1} \\
  &+ (s_2 + \alpha s_1 + \gamma_+ k_2 + \gamma_- k_1)\partial_{k_2}\\
  &- (k_2+ \beta k_1 - \gamma_+ s_2 - \gamma_- s_1)\partial_{s_2}\\
  &-{1\over 2} \, \vec{R}^{T} \, \mathrm{M}_D\,  \vec{R}\,,
  \end{aligned}
\end{equation}
with $\gamma_{\pm} = (\gamma_2 \pm \gamma_1)/4$ and 
$\mathrm{M}_D$ being the diffusion matrix of the global correlated dissipative environment given by:
\begin{equation}\label{MDglobal}
\mathrm{M}_D = {1\over 4}\left(
  \begin{array}{cccc}
     d_1^{(k)}+ d_2^{(k)} & 0 & d_2^{(k)}- d_1^{(k)} & 0\\
     0 & d_1^{(s)}+ d_2^{(s)} & 0 & d_2^{(s)}- d_1^{(s)}\\
     d_2^{(k)}- d_1^{(k)} & 0 & d_1^{(k)}+ d_2^{(k)} & 0\\
     0 & d_2^{(s)}- d_1^{(s)} & 0 & d_1^{(s)}+ d_2^{(s)}
   \end{array}\right)\,,
\end{equation}
where
\begin{equation}
d_j^{(k)} \equiv \gamma_j N_j\sqrt{{\lambda_j(\beta)\over \lambda_j(\alpha)}}, 
\quad d_j^{(s)} \equiv \gamma_j N_j \sqrt{{\lambda_j(\alpha)\over \lambda_j(\beta)}}, 
\end{equation}
and 
\begin{equation}
N_j\equiv 2\bar n_j+1,\quad \lambda_1(y)\equiv 1-y, \quad \lambda_2(y) \equiv 1 + y.
\end{equation}

Regardless of the type of dissipative environment, all the partial differential equations are
particular cases of a generic partial differential equation:
\begin{eqnarray}\nonumber
\left( \mathcal{L}_o + 2 \mu \partial_{k_1} + 2 \nu  \partial_{k_2} \right) w(\vec{R}, t)
&=& \rmi ( \Delta + \vec{\delta}_{\xi,\zeta}\cdot\vec{R})\, w(\vec{R}, t),\\
\end{eqnarray}
where $\vec{\delta}_{\xi,\zeta} = (0,\xi,0,\zeta)^T$.
This PDE of the first order can be solved by employing the method of characteristics. 
The liner partial differential equation can be placed in the 
parametric form:
 \begin{eqnarray}\label{parametricPDE}
{\rmd\over\rmd t} \vec{R} (t) &=& {\bf A} \vec{R}(t) + 2\vec{\eta}_{\mu,\nu},\\
{\rmd \over \rmd t} w(\vec{R}, t) &=& \rmi ( \Delta + \vec{\delta}_{\xi,\zeta} \cdot\vec{R}\,) w(\vec{R}, t)\\
&&-{1\over 2} \, \left( \vec{R}^{T} \, \mathrm{M}_D\,  \vec{R} \right)\, w(\vec{R}, t) \,,
 \end{eqnarray}
 where ${\bf A}$ is the stability matrix of the two-oscillator system, which for the 
 local dissipation case, it reads:
 \begin{equation}\label{disstabilityMat}
{\bf A} = \left(
  \begin{array}{cccc}
     \gamma_1 & 1 & 0 & \alpha\\
     -1 & \gamma_1 & -\beta & 0\\
     0 & \alpha & \gamma_2 & \Omega\\
     -\beta & 0 & -\Omega & \gamma_2
   \end{array}\right)\,,
 \end{equation}
while for the global correlated dissipative environment, the stability matrix takes a different form:
 \begin{equation}
{\bf A} = \left(
  \begin{array}{cccc}
     \gamma_+ & 1 & \gamma_- & \alpha\\
     -1 & \gamma_+ & -\beta & \gamma_-\\
     \gamma_- & \alpha & \gamma_+ & 1\\
     -\beta & \gamma_- & -1 & \gamma_+
   \end{array}\right)\,.
 \end{equation}
Additionally in Eq. (\ref{parametricPDE}) we have defined $\vec{\eta}_{\mu,\nu}  \equiv (\mu, 0 , \nu, 0)^{T}$. 
The solution for the first of these ordinary differential equations 
is directly obtained:
 \begin{equation}\nonumber
   \vec{R}(t) = \bPh(t-t_o)\vec{R}(t_o) + 2 \vec{\eta}_{\mu,\nu}(t-t_o),
 \end{equation}
 where
 \begin{equation}\label{xiap}
   \vec{\eta}_{\mu,\nu}(t-t_o)   =  \int_{t_o}^{t}\rmd t' \bPh(t-t') \vec{\eta}_{\mu,\nu}.
\end{equation}
This solution has been obtained in terms of the transition
matrix $\bPh$ which is nothing but the exponentiation of the stability
matrix ${\bf A}$:
\begin{equation}
 \bPh(t) = \exp({\bf A} \, t),
\end{equation}
fulfilling the group properties, \ie:
$\bPh(t + s) =\bPh(t)\bPh(s)$,
$\bPh(t=0) =\One$, $\bPh^{-1}(t) =\bPh(-t)$.

The solution for the second equation can be derived through the employment of the
transition matrix because of its group properties; thus by noticing that
\begin{equation}\label{inmap}
   \vec{R}(t') = \bPh(t'-t)\vec{R}(t) + 2\vec{\eta}_{\mu,\nu}(t'-t),
\end{equation}
where $\bPh(t'-t) = \bPh^{-1}(t-t')$ and
\begin{eqnarray}\label{etavec}
  \vec{\eta}_{\mu,\nu}(t'-t) &=& - \bPh^{-1}(t-t')\vec{\eta}_{\mu,\nu}(t-t')\\\nonumber
  &=& - \int_{t'}^t\rmd t'' \bPh(t'-t'')\vec{\eta}_{\mu,\nu},
\end{eqnarray}
then integration of the second equation can be formulated as:
\begin{eqnarray}\nonumber
\int_{w(t_o)}^{w(t)}  {\rmd w \over w} &=& \rmi \int_{t_o}^{t} \rmd t'
 \left\{ \Delta + \vec{\delta}_{\xi,\zeta}\cdot\vec{R}(t')\right\}\\
 &&-{1\over 2} \int_{t_o}^{t} \rmd t' \,\vec{R}^{T}(t') \, \mathrm{M}_D\,  \vec{R}(t')  ,
\end{eqnarray}
and by employing Eq. (\ref{inmap}):
\begin{eqnarray}
\int_{w(t_o)}^{w(t)}  {\rmd w \over w} &=& \rmi \Delta \cdot (t-t_o)\\\nonumber
&+&\rmi\!\! \int_{t_o}^{t} \!\!\!\!\rmd t'
\vec{\delta}_{\xi,\zeta}\cdot(\bPh(t'-t)\vec{R}(t) +
2\vec{\eta}_{\mu,\nu}(t'-t))\\\nonumber
&-& {1\over 2} \int_{t_o}^{t} \rmd t'\left[ (\bPh(t'-t)\vec{R}(t) +
2\vec{\eta}_{\mu,\nu}(t'-t))^{T} \, \mathrm{M}_D\right. \\\nonumber
&&\hspace{1.5cm} \left.(\bPh(t'-t)\vec{R}(t) +
2\vec{\eta}_{\mu,\nu}(t'-t))\right] ,
\end{eqnarray}
Integration of the involved terms can be easily carried out using the properties of the
transition matrix. Two of the most relevant integrals involves the transition matrix 
$\bPh(t)$ or its inverse $\bPh(-t)$ which are easily obtained by noticing that the transition matrix 
is the solution of the following ordinary differential equation:
\begin{equation}
  {\rmd\over \rmd t} \bPh(t) = \bA \bPh(t),\quad {\rmd\over \rmd t} \bPh^{-1}(t) = -\bPh^{-1}(t) \bA, 
\end{equation}
thus integrating in both sides and solving for the involved integrals yields:
\begin{eqnarray}
  \int_{t_o}^t \rmd t'\bPh(t') &=& \bA^{-1} \left(\bPh(t)- \bPh(t_o) \right), \\
  \int_{t_o}^t \rmd t' \bPh^{-1}(t') &=& - \left(\bPh^{-1}(t) - \bPh^{-1}(t_o) \right)\; \bA^{-1}. 
\end{eqnarray}
Notice that in the case when: $\bPh(t_o=0) = \bPh^{-1}(t_o=0) =\One$.
The solution of the first order partial differential equation is therefore given as:
\begin{eqnarray}
  w (\vec{R},t) &=& w (\bPh^{-1}(t-t_o)\vec{R} + 2\vec{\eta}_{\mu,\nu}(t_o-t),t_o)\\\nonumber
  && \exp\left( \rmi \Delta\, (t-t_o)  -2 \rmi \Gamma^{\xi,\zeta}_{\mu,\nu}(t-t_o) - 2\varphi_{\mu,\nu}(t-t_o)\right)\\\nonumber
  && \exp\left( \left[\rmi \vec{\delta}_{\xi,\zeta}(t-t_o) - 2\vec{\chi}_{\mu,\nu}(t-t_o)\right] \cdot \vec{R} \right) \\\nonumber
  && \exp\left( -{1\over 2} \vec{R}^{T}\,\mathrm{M}(t-t_o)\, \vec{R} \right) ,  
\end{eqnarray}
where 
\begin{equation}\label{deltavec}
  \vec{\delta}_{\xi,\zeta}(t-t_o) = \int_{t_o}^{t}\rmd t' \bPh^{T}(t'-t) \vec{\delta}_{\xi,\zeta},
\end{equation}
\begin{equation}
  \Gamma^{\xi,\zeta}_{\mu,\nu}(t-t_o) = \int_{t_o}^{t} \rmd t'\, \int_{t'}^{t} \rmd t''\, 
  \bPh^{T}(t'-t'')\, 
  \vec{\delta}_{\xi,\zeta} \cdot \vec{\eta}_{\mu,\nu},
\end{equation}
\begin{equation}
  \mathrm{M}(t-t_o) = \int_{t_o}^{t} \rmd t'\, 
  \bPh^{T}(t'-t)\, \mathrm{M}_D\, \bPh(t'-t), 
\end{equation}
\begin{equation}
  \vec{\chi}_{\mu,\nu}(t-t_o) = \int_{t_o}^{t} \rmd t'\, \bPh^{T}(t'-t)\, \mathrm{M}_D \,\vec{\eta}_{\mu,\nu}(t'-t)
\end{equation}
\begin{equation}
  \varphi_{\mu,\nu}(t-t_o) = \int_{t_o}^{t} \rmd t'\, 
  \vec{\eta}^{T}_{\mu,\nu}(t'-t)\, \mathrm{M}_D\, \vec{\eta}_{\mu,\nu}(t'-t), 
\end{equation}
and we have made the substitution of the 
initial condition coordinates $\vec{R}(t_o) \rightarrow \vec{R} = \vec{R}(t)$, 
through the map given at (\ref{inmap}), \ie: 
\[ w (\vec{R}(t_o),t_o) \rightarrow w (\bPh^{-1}(t-t_o)\vec{R} + 2\vec{\eta}_{\mu,\nu}(t_o-t),t_o).\] 
With this general solution, each of the elements of the system of partial differential 
equations given in (\ref{rho11})-(\ref{rho44}) can be obtained by just substituting the 
corresponding values of $\Delta$, $\xi$, $\zeta$, $\mu$ and $\nu$ for each element. 
Particular cases depend upon the initial conditions, which are ultimately encoded 
in the $c_{ij}$ coefficients and the characteristic function $w(\vec{R}(t_o))\equiv w(\vec{R}_o,t_o)$.

We will focus 
only on an initial preparation of the two-qubit subsystem in a Bell state.
This configuration is achieved by applying a projective 
operator $\h{P}=\One_{o's} \otimes |\Psi^+\rangle\langle\Psi^+|$ to the initially 
decoupled configuration (\ref{ic}).
The projection operator, written in the chosen computational basis, reads: 
$\h{P}=\One_{o's} \otimes {1\over 2}(|2\rangle\langle 2 | + |3\rangle\langle 3| + 
|3 \rangle\langle 2| + |2 \rangle\langle 3|)$. 
This yields the following initial state for the full composite system:
\begin{eqnarray}\label{Bellw}
  \tilde{\varrho}(\vec{R}_o, t_o) &=&
  \h{P}\varrho(\vec{R}_o,t_o)\h{P}^{\dag}\\\nonumber
  &=& \left(\begin{array}{cccc}
    0 & 0 & 0 & 0\\
    0 & A & A & 0\\
    0 & A & A & 0\\
    0 & 0 & 0 & 0
    \end{array} \right)\,w(\vec{R}_o,t_o)\,.
\end{eqnarray}
where $A =(c_{22} + c_{33} + c_{23} + c_{32})/2$. In this regard, the initial Bell state of the two-qubit subsystem 
yields only the elements 
$w_{22}$, $w_{23}$, $w_{32}$, $w_{33}$ different from zero in the solution of the full composite system. 
For these elements, the values of $\Delta$, $\xi$, $\zeta$, $\mu$ and $\nu$ are given in the following table:
\begin{center}
\begin{tabular}{c|c|c|c|c|c}
              & $\Delta$ & $\xi$ & $\zeta$ & $\mu$ & $\nu$ \\
  \hline
  $w_{22}$ & $0$ & $g_1$ & $-g_2$ & $0$ & $0$ \\
  $w_{23}$ & $\Delta_1 - \Delta_2$ & 0 & 0 & $-g_1$ & $g_2$ \\
  $w_{32}$ & $\Delta_2 - \Delta_1$ & 0 & 0 & $g_1$ & $-g_2$ \\ 
  $w_{33}$ & $0$ & $-g_1$ & $g_2$ & $0$ & $0$ \\
\end{tabular}
\end{center}
On the other hand, to track concurrence in the two-qubit subsystem, the oscillator degrees of freedom 
must be traced out, which is achieved by evaluating the total state at $\vec{R}=0$. The evolution to a time $t$ from the 
initial time $t_o=0$ of the reduced 
two-qubit state is then given by:
\begin{eqnarray}\label{solentq}
    \tilde{\varrho}_{\text{q's}}(t) &=&  {1\over 2} \left(\begin{array}{cccc}
      0 & 0 & 0 & 0\\
      0 & 1 &   f(t) & 0\\
      0 &  f^{*}( t)  & 1 & 0\\
      0 & 0 & 0 & 0
      \end{array} \right)\,, 
\end{eqnarray}
where
\begin{equation}\label{ft}
  f( t) = w\left( 2\vec{\eta}_{-g_1,g_2}(-t)\,\right)\,\rme^{\rmi (\Delta_1-\Delta_2) \, t - 2 \varphi_{-g_1,g_2}(t-t_o)} \,.
\end{equation}

\section{\label{app3}
Derivation of the master equation with global dissipation and its characteristic-function
representation}

In this appendix, we derive the global master equation describing dissipation in the interacting
two-oscillator subsystem. The derivation is performed in the normal-mode basis of the
coupled oscillators, obtained through a canonical Bogoliubov transformation
\cite{Bogoliubov1958,Valatin1958, Zhang2015, breuer2002}.

We focus on the resonant configuration $\Omega=1$, for which the oscillator Hamiltonian can be 
diagonalized using an orthogonal transformation followed by a mode-dependent canonical rescaling.
Although the procedure can be extended to nonresonant oscillators, the resonant case allows a 
particularly simple analytical treatment.

The dissipative description is constructed in the normal-mode basis of the interacting oscillator 
subsystem. Accordingly, the resulting master equation incorporates the oscillator--oscillator interaction
into the definition of the dissipative operators. We restrict the derivation to parameter configurations 
for which the quadratic oscillator Hamiltonian is bounded from below and the normal modes possess positive 
energies.

\subsection*{Normal-mode representation}

The Hamiltonian of the interacting oscillators is
\begin{eqnarray}\nonumber
H_{\mathrm{osc's}}
&=&
\h{n}_a+\h{n}_b + \alpha\h{x}_a\h{x}_b + \beta\h{p}_a\h{p}_b\\
&=&
\frac12\left( \vec{p}^{\,T}\mathrm M(\beta)\vec{p}
+ \vec{x}^{\,T}\mathrm M(\alpha)\vec{x} \right),
\end{eqnarray}
where $\vec{x}=(\h{x}_a,\h{x}_b)^T$, $\vec{p}=(\h{p}_a,\h{p}_b)^T$,
and $\h{n}_j=\h{a}_j^\dagger\h{a}_j$. The zero-point energies are included
through the identity $\h{n}_j+1/2 = (\h{x}_j^2+\h{p}_j^2)/2$.
Thus, the first equality above is understood up to the irrelevant
constant zero-point contribution.
The matrices $\mathrm M(y)$, with $y=\alpha,\beta$, are
\begin{equation}
\mathrm M(y)=
\begin{pmatrix}
1&y\\
y&1
\end{pmatrix}
\end{equation}
which in turn can be simultaneously diagonalized by the 
orthogonal involutory transformation
\begin{equation}
\mathrm S=\frac1{\sqrt2}
\begin{pmatrix}
-1&1\\
1&1
\end{pmatrix}, \qquad \mathrm S^{-1}=\mathrm S,
\end{equation}
such that
\begin{equation}
\mathrm M(y)= \mathrm S\Lambda(y)\mathrm S,
\end{equation}
with
\begin{equation}
\Lambda(y)=
\mathrm{diag}(\lambda_1(y),\lambda_2(y)) = \mathrm{diag}(1-y,1+y).
\end{equation}

Introducing the normal-mode quadratures
\begin{equation}\label{pm_normal_modes}
\vec X=\mathrm S\vec x, \qquad \vec P=\mathrm S\vec p,
\end{equation}
the two-oscillator Hamiltonian becomes
\begin{eqnarray}\nonumber
H_{\mathrm{osc's}}
&=& \frac12\left[
\vec P^{\,T}\Lambda(\beta)\vec P + \vec X^{\,T}\Lambda(\alpha)\vec X
\right]\\
&=&
\frac12\sum_{k=1}^{2} \left[ \lambda_k(\beta)P_k^2 + \lambda_k(\alpha)X_k^2
\right].
\end{eqnarray}
whose normal-mode frequencies are given as\
\begin{equation}
\tilde\omega_k= \sqrt{\lambda_k(\alpha)\lambda_k(\beta)}.
\end{equation}

In order to guarantee positive-energy normal modes, we impose
\begin{equation}\label{global_stability_condition}
\lambda_k(\alpha)>0, \qquad \lambda_k(\beta)>0, \qquad k=1,2.
\end{equation}
For $\Omega=1$, these conditions are equivalent to
\begin{equation}
|\alpha|<1,\qquad |\beta|<1.
\end{equation}

They ensure both real positive normal-mode frequencies and a quadratic Hamiltonian 
bounded from below. These conditions are more restrictive than requiring real 
frequencies alone, \ie $\lambda_k(\alpha)\lambda_k(\beta)>0$.

The normal-mode operators are defined by
\begin{eqnarray}\nonumber
A_k
&=&
\frac1{\sqrt2}\left(\frac{\lambda_k(\alpha)}{\lambda_k(\beta)}\right)^{1/4}X_k 
+\frac{\rmi}{\sqrt2}\left(\frac{\lambda_k(\beta)}{\lambda_k(\alpha)}\right)^{1/4}P_k\\
&=&
\sum_{j=a,b}
\left( u_{kj}\h{a}_j + v_{kj}\h{a}_j^\dagger \right),
\end{eqnarray}
where
\begin{eqnarray}
u_{kj}
&=&
\frac12\left[\left(\frac{\lambda_k(\alpha)}{\lambda_k(\beta)}\right)^{1/4}
+\left(\frac{\lambda_k(\beta)}{\lambda_k(\alpha)}\right)^{1/4} \right]S_{kj},\\
v_{kj}
&=&
\frac12 \left[\left( \frac{\lambda_k(\alpha)} {\lambda_k(\beta)} \right)^{1/4}
- \left( \frac{\lambda_k(\beta)} {\lambda_k(\alpha)} \right)^{1/4} \right]S_{kj}.
\end{eqnarray}

Within the parameter region (\ref{global_stability_condition}),
the Bogoliubov coefficients are real and the operators satisfy
$ [A_k,A_l^\dagger]=\delta_{kl}$. Consequently,
\begin{equation}\label{Hosc_global_diag}
H_{\mathrm{osc's}} =
\sum_{k=1}^{2} \tilde\omega_k \left(A_k^\dagger A_k+\frac12\right).
\end{equation}

\subsection*{Global master equation}

We now introduce a thermal environment coupled directly to the normal modes
of the interacting oscillator subsystem.
The total Hamiltonian is
\begin{equation}
H_{\mathrm{total}}
= H_{\mathrm{o}}+H_{\mathrm{env}} + H_{\mathrm{int}},
\end{equation}
where
\begin{equation}
H_{\mathrm{o}} = H_{\mathrm{qubits}} + H_{\mathrm{coupling}}
+ H_{\mathrm{osc's}},
\end{equation}
and
\begin{equation}
H_{\mathrm{env}} = \sum_l\nu_l\h{b}_l^\dagger\h{b}_l.
\end{equation}

Following the rotating-wave approximation for the oscillator--bath interaction, 
we consider
\begin{eqnarray}\nonumber
H_{\mathrm{int}}
&=& \sum_{k=1}^{2}\sum_l \left( h_l A_k^\dagger\h{b}_l + h_l^* A_k\h{b}_l^\dagger \right)\\
&=& \sum_{k=1}^{2} \left( A_k B^\dagger + A_k^\dagger B \right),
\end{eqnarray}
where $B=\sum_lh_l\h{b}_l$. The environment is assumed to be initially in a thermal 
equilibrium state, $\varrho_{\mathrm{env}}
=\exp(-H_{\mathrm{env}}/T) / \tr[\exp(-H_{\mathrm{env}}/T)]$, where $k_B=1$, and 
we assume initially factorized system--environment conditions, weak oscillator--bath 
coupling, and a bath correlation time much shorter than the dissipative relaxation 
times. 

The derivation of the dissipative generator is performed using the 
oscillator normal modes of Eq.~(\ref{Hosc_global_diag}) as the reference transition 
operators, \ie in constructing the bath-induced dissipator, we use the interaction picture 
associated with $H_{\mathrm{osc's}}+H_{\mathrm{env}}$, while retaining the qubit--oscillator
interaction in the coherent part of the master equation; in this context, 
the system--environment interaction in the interaction picture is
\begin{eqnarray}
\tilde H_{\mathrm{int}}(t) = \sum_{k=1}^{2} \left[ A_k e^{-\rmi\tilde\omega_k t} B^\dagger(t) 
+ A_k^\dagger e^{\rmi\tilde\omega_k t} B(t) \right].
\end{eqnarray}

This constitutes an oscillator-normal-mode approximation to the dissipative dynamics, 
which is justified when dissipation is targeted indirectly at the subsystem of 
interest~\cite{Gorin2016,Lopez2020}, and is especially well justified in the dephasing-coupling scenario, 
for which transitions in the energy spectrum are suppressed. In other words, the main mechanism 
of dissipation acts on the oscillator modes that couple significantly to the environment, while 
the qubit degrees of freedom are only indirectly affected. This may occur because the large 
coupling strengths to the two-oscillator system shield the qubits from direct environmental 
influence, because their transition frequencies are far detuned from the environmental 
spectrum, or because the system has been specifically engineered so that the qubits are 
affected only indirectly.

On the other hand, a fully dressed global derivation based on the complete Hamiltonian $H_{\mathrm{o}}$ 
would additionally account for qubit-dependent displacements of the oscillator modes which, depending 
on the strength of the qubit--bath and qubit--oscillator couplings, could modify the dissipative dynamics, 
although for the parameter regime considered here, such modifications are expected to be perturbative.
In this context, the oscillator-normal-mode approximation provides a reliable framework for deriving 
the effective dissipative dynamics of the two-oscillator subsystem, while the qubits remain largely 
unaffected by direct environmental interactions.  

The standard approximations for deriving the master equation 
first consist on performing a perturbative expansion of the von-Neumann equation in the interaction picture 
for the total system up to second order in the system-environment interaction Hamiltonian, tracing over the 
environmental degrees of freedom and afterwards making the Born approximation which assumes that the 
environment remains in its initial state throughout the evolution; secondly, the Markov approximation is 
applied which assumes that the system's memory effects are negligible and that the bath correlation functions 
decay rapidly compared to the system's evolution timescale. Quantitatively, the Markov approximations consists 
of considering the density operator of the system as approximately constant over the timescale dynamics of the 
bath correlations while extending the upper integration limit to infinity. Under these approximations, the resulting 
master equation takes the form of a time-local differential equation for the reduced density operator:
\begin{equation}\label{Markov}
{\rmd \over \rmd t} {\tilde{\varrho}}(t) 
= - \int_0^{t} \rmd t' \, 
\tr_{\mathrm{env}} \left[\tilde{H}_{\mathrm{int}}(t), [\tilde{H}_{\mathrm{int}}(t'), 
\tilde{\varrho}(t) \otimes \varrho_{\mathrm{env}}]\right],
\end{equation}
where $\tilde{\varrho}(t)$ denotes the reduced density operator of the two-qubit two-oscillator 
subsystem in the interaction picture.

For a common environment, the resulting generator generally contains contributions connecting 
different normal modes, with oscillating factors
\begin{equation}
\exp\left[\rmi(\tilde\omega_k-\tilde\omega_{k'})t\right];
\end{equation}
these terms may be neglected under the secular approximation when the normal-mode frequencies are
sufficiently separated relative to the dissipative relaxation and cross-coupling rates,
\begin{equation}\label{global_secular_condition}
|\tilde\omega_1-\tilde\omega_2| \gg \gamma_1,\gamma_2,
\end{equation}
up to the relevant mode-dependent coupling factors. In this context, for exactly degenerate normal modes, 
the cross-frequency terms do not oscillate and must generally be retained. In particular, for 
$\alpha=\beta=0$ and $\Omega=1$, the frequencies satisfy $\tilde\omega_1=\tilde\omega_2=1$. 
Therefore, a single common reservoir cannot generally be represented by independent normal-mode 
dissipators at this point.

In the present work, the global description is applied to interacting-oscillator configurations
satisfying the spectral-separation condition (\ref{global_secular_condition}).
The uncoupled configuration is instead represented through the local dissipative model, providing
a reference corresponding to a distinct reservoir description.

The bath correlation functions are defined as $C_{\downarrow}(\tau) =
\tr_{\mathrm{env}}\left[B(\tau)B^\dagger(0)\varrho_{\mathrm{env}} \right]$, and 
$C_{\uparrow}(\tau) = \tr_{\mathrm{env}}\left[B^\dagger(\tau)B(0)\varrho_{\mathrm{env}}\right]$;
then by introducing a common form for the spectral density functions: 
$J(\nu) = \sum_l|h_l|^2\delta(\nu-\nu_l)$, and a thermal occupation: $\bar n(\nu)=(\exp(\nu/T)-1)^{-1}$, 
the bath correlation functions take the form:
\begin{eqnarray}
C_{\downarrow}(\tau)
&=& \int_0^\infty\rmd\nu\, J(\nu)[\bar n(\nu)+1] e^{-\rmi\nu\tau},\\
C_{\uparrow}(\tau)
&=& \int_0^\infty\rmd\nu\, J(\nu)\bar n(\nu) e^{\rmi\nu\tau}.
\end{eqnarray}

Performing the bath integrations, using the Sokhotski--Plemelj identity~\cite{Vladimirov1971}, 
and retaining the secular contributions, we obtain
\begin{eqnarray}\label{GME1}
{\rmd \over \rmd t} {\varrho}
&=& -\rmi [H_{\mathrm{o}} + H_{\mathrm{LS}},\varrho ]\\\nonumber
&& + \sum_{k=1,2} \Big[ \gamma_k (\bar{n}(\tilde{\omega}_k) + 1) \mathcal{D}[A_k] \varrho 
+ \gamma_k \bar{n}(\tilde{\omega}_k) \mathcal{D}[A_k^{\dagger}] \varrho \Big]
\end{eqnarray}
where $\gamma_k=2\pi J(\tilde\omega_k)$,
$\bar n_k=\bar n(\tilde\omega_k)$, and $\mathcal D[O]\varrho$ represents the most relevant 
algebraical structure in the literature about possitive preserving quantum dynamical 
semigroups:
\begin{equation}
\mathcal D[O]\varrho =
O\varrho O^\dagger - \frac12\{O^\dagger O,\varrho\}.
\end{equation}
The coefficients in the global master equation $\gamma_k$, which depend on the bath spectral 
density and the normal-mode frequencies, may differ from those of a local description, because 
for the local description, the reservoir spectral density is evaluated at the corresponding local 
oscillator frequencies. Nevertheless, these differences are typically small under the 
weak-coupling assumptions and could be treated perturbatively.

The Lamb-shift Hamiltonian is
\begin{equation}\label{HLS_global}
H_{\mathrm{LS}}
= \sum_{k=1}^{2} \delta_k A_k^\dagger A_k + \mathrm{constant},
\end{equation}
where, for the rotating-wave oscillator--bath coupling considered,
\begin{equation}\label{delta_LS_global}
\delta_k= \mathcal P\int_0^\infty \rmd\nu\, \frac{J(\nu)}{\tilde\omega_k-\nu}.
\end{equation}
were $\mathcal P$ denotes the Cauchy principal value. Notice that
the Lamb-shift coefficients also depend on the bath spectral density and its ultraviolet 
regularization (the cutoff frequency) and can also be considered as perturbative corrections 
under the weak-coupling assumptions used in the derivation.

Using the normal-mode quadratures, Eq.~(\ref{HLS_global}) can be expressed, up to an additive 
constant, as
\begin{equation}\label{HLS_quadratures}
H_{\mathrm{LS}}
=
\frac12\sum_{k=1}^{2}
\delta_k \left( q_kX_k^2+q_k^{-1}P_k^2 \right),
\end{equation}
with
\begin{equation}
q_k= \sqrt{ \frac{\lambda_k(\alpha)}{\lambda_k(\beta)}}.
\end{equation}

Thus, the Lamb shift preserves the quadratic character of the oscillator Hamiltonian and
renormalizes its normal-mode frequencies. When transformed back to the original oscillator
variables, these corrections modify both the local quadratic terms and the oscillator--oscillator
interaction coefficients.
Nevertheless, since the Lamb-shift corrections are perturbative under the adopted
weak-coupling assumptions, and our objective is to investigate the role of the tunable oscillator
interactions, we neglect $H_{\mathrm{LS}}$ in the numerical results presented in the main text.
This approximation assumes that the associated frequency shifts remain small compared with the
relevant dynamical scales and do not drive the effective Hamiltonian across a stability
boundary.

The global dissipator then takes the form
\begin{equation}
\mathcal D_{\mathrm{global}}[\varrho]
= \sum_{k=1}^{2} \gamma_k
\left[ (\bar n_k+1)\mathcal D[A_k]\varrho
+\bar n_k\mathcal D[A_k^\dagger]\varrho \right].
\end{equation}

\subsection*{Position--momentum representation}

We now transform the dissipative contribution into the normal-mode
position--momentum representation. Using the definitions of the
operators $A_k$, the dissipator can be written as
\begin{equation}
\mathcal D_{\mathrm{global}}[\varrho]
= \mathcal L_1[\varrho] + \mathcal L_2[\varrho],
\end{equation}
where
\begin{equation}
\begin{aligned}
\mathcal L_k[\rho]
={}&
-\frac{i\gamma_k}{4}
\Big(
[X_k,\{P_k,\rho\}]
- [P_k,\{X_k,\rho\}]
\Big)
\\
&-
\frac{\gamma_kN_k}{4}
\Big(q_k[X_k,[X_k,\rho]] +
q_k^{-1}[P_k,[P_k,\rho]]
\Big),
\end{aligned}
\end{equation}
with
\begin{equation}
N_k=2\bar n_k+1,
\qquad
q_k=
\sqrt{
\frac{\lambda_k(\alpha)}
{\lambda_k(\beta)}
}.
\end{equation}

The normal-mode quadratures are related to the original variables by
\begin{equation}
X_k=\sum_{j=a,b}S_{kj}x_j,
\qquad
P_k=\sum_{j=a,b}S_{kj}p_j.
\end{equation}

The first contribution to $\mathcal L_k$ describes damping, whereas the double-commutator
terms describe thermal diffusion. Although the modes are independently damped in the 
normal-mode basis, their transformation into the original oscillator coordinates
produces correlated dissipative contributions. For $\alpha\neq\beta$, the Bogoliubov 
transformation additionally introduces anomalous terms associated with the squeezing structure of
the normal-mode representation. These terms originate from ordinary thermal damping of the 
collective modes and do not, by themselves, imply that the external reservoir is in a squeezed 
thermal state.

\subsection*{Characteristic-function representation}

We introduce the normal-mode Fourier phase-space variables
\begin{equation}
\kappa_k=
\sum_jS_{kj}k_j,
\qquad
\sigma_k=
\sum_jS_{kj}s_j,
\end{equation}
and the associated derivatives
\begin{equation}
\partial_{\kappa_k}
=
\sum_jS_{kj}\partial_{k_j},
\qquad
\partial_{\sigma_k}
=
\sum_jS_{kj}\partial_{s_j}.
\end{equation}

Applying the characteristic-function transformation rules given in Eqs.~(\ref{xpr}--\ref{prx}),
the global dissipator becomes
\begin{equation}\label{DGME1}
\begin{aligned}
\mathcal D_{\mathrm{global}}w
={}&
-\sum_{k=1}^{2}
\Bigg[
\frac{\gamma_k}{2}
\left(
\kappa_k\partial_{\kappa_k}
+
\sigma_k\partial_{\sigma_k}
\right)\\
&\qquad+
\frac{\gamma_kN_k}{4}
\left(
q_k^{-1}\kappa_k^2+
q_k\sigma_k^2
\right)
\Bigg]w.
\end{aligned}
\end{equation}

Consequently, the dissipative contribution retains thefirst-order drift 
and quadratic diffusion structure of a Gaussian characteristic-function
equation. Transforming back to the original Fourier phase-space variables,
$\vec R=(k_1,s_1,k_2,s_2)^T$, we obtain
\begin{equation}
\begin{aligned}
\mathcal D_{\mathrm{global}}
w(\vec R,t)
={}&
\vec f(\vec R)\cdot
\nabla_{\vec R}w(\vec R,t)\\
&-\frac12
\vec R^{\,T}\mathrm M_D\vec R\,
w(\vec R,t),
\end{aligned}
\end{equation}
where
\begin{equation}
\vec f(\vec R)=-
\begin{pmatrix}
\gamma_+ k_1+\gamma_- k_2\\
\gamma_+ s_1+\gamma_- s_2\\
\gamma_- k_1+\gamma_+ k_2\\
\gamma_- s_1+\gamma_+ s_2
\end{pmatrix},
\end{equation}
and
\begin{equation}
\gamma_\pm=
\frac{\gamma_2\pm\gamma_1}{4}.
\end{equation}

The diffusion matrix is
\begin{equation}\label{MDglobal}
\mathrm M_D=
\frac14
\begin{pmatrix}
d_1^{(k)}+d_2^{(k)}&0&
d_2^{(k)}-d_1^{(k)}&0\\
0&d_1^{(s)}+d_2^{(s)}&0&
d_2^{(s)}-d_1^{(s)}\\
d_2^{(k)}-d_1^{(k)}&0&
d_1^{(k)}+d_2^{(k)}&0\\
0&d_2^{(s)}-d_1^{(s)}&0&
d_1^{(s)}+d_2^{(s)}
\end{pmatrix},
\end{equation}
where
\begin{equation}
d_j^{(k)}
=\gamma_jN_jq_j^{-1},
\qquad
d_j^{(s)}
=\gamma_jN_jq_j.
\end{equation}

The resulting expressions coincide with the drift and diffusion structure 
employed in the analytical characteristic-function solutions of the global 
dissipative model. In particular, the global dissipator produces no derivatives
higher than first order and preserves the Gaussian structure of the dynamical 
equations. Its effects are completely encoded in the dissipative drift vector 
and the diffusion matrix $\mathrm M_D$.

If the Lamb shift is retained, its contribution is incorporated through an additional 
quadratic Hamiltonian drift, without modifying the differential order of the 
characteristic-function equations.

\begin{acknowledgments}
J.R.L. acknowledges a doctoral fellowship from SECIHTI. 
\end{acknowledgments}

\noindent \textbf{Data Availability Statement:} 
The data used for the generation of the figures of this article were produced by the results 
given in the main text or the appendices of the article.

\bibliography{man}   % name your BibTeX data base

\end{document}